# Cognitive Alignment Drives Attention: Modeling and Supporting Socially Shared Regulation in Pair Programming

Anahita Golrang, Kshitij Sharma

Department of Computer Science, Norwegian University of Science and Technology

Anahita.m.golrang@ntnu.no

**Abstract**: Effective collaboration in computer-supported collaborative learning (CSCL) depends not only on task structure or technological affordances, but on learners' ability to regulate cognition and attention together over time. Grounded in the theory of socially shared regulation of learning (SSRL), this paper investigates how joint mental effort (JME) and joint visual attention (JVA) function as process-level indicators of shared regulation in pair programming, and how AI-driven adaptive feedback can strengthen these regulatory processes.

We present three eye-tracking studies involving 182 dyads engaged in collaborative debugging tasks. Study 1 examines natural collaboration and shows that high-performing dyads exhibit significantly higher JME and JVA, a greater prevalence of productive high-JME–high-JVA episodes, and a stable causal relationship in which JME predicts JVA. Study 2 evaluates reactive adaptive feedback based on real-time deviations in JME and/or JVA. Results show that combined feedback targeting both dimensions yields the strongest improvements in performance, regulatory coherence, and cognitive-to-attentional causality, outperforming single-channel feedback. Study 3 introduces proactive, forecast-based feedback using machine-learning predictions of future collaboration states. Proactive support further enhances performance and sustains shared regulation by anticipating breakdowns before they manifest.

Across studies, causal modeling reveals that cognitive alignment systematically drives attentional coordination in successful collaboration, while mismatches between effort and attention characterize unproductive regulation. Methodologically, this work integrates dual eye-tracking, pupillometry, episode-based analysis, and causal inference to capture SSRL as a dynamic, emergent process. Conceptually, the findings position AI not as an automated controller, but as an intelligence-augmenting co-regulator that supports learners' capacity to coordinate effort, attention, and understanding together.

**Keywords**: Adaptive collaborative learning support, Computer-supported collaborative learning (CSCL), Dual eye-tracking, eye-tracking, joint mental effort, joint visual attention, Multimodal learning analytics, pair programming

# Introduction and Background: Role of AI in CSCL

Computer-Supported Collaborative Learning (CSCL) has evolved into a rich, interdisciplinary research field concerned with how technological, pedagogical, and social factors interact to shape collaborative learning processes and outcomes. Over the past decades, research has progressively moved from foundational questions about structuring collaboration toward more sophisticated approaches that model learners and groups, analyze interaction data at scale, and provide adaptive, intelligent support. The use of AI in CSCL research reflects a growing recognition that effective collaboration is not solely a matter of task design or technology provision, but rather emerges from the dynamic interplay between instructional structure, shared artifacts, learner characteristics, affective and motivational processes, and real-time regulation.

A central theme in CSCL research is the role of instructional structure in shaping collaboration quality. Numerous studies show that scripted collaboration, through predefined roles, prompts, or interaction sequences, leads to more elaborated argumentation, increased engagement, and higher-quality discourse compared to unstructured collaboration (Scheuer et al, 2011; Lambropoulos & Bratitsis, 2014). Scripts are particularly effective for supporting learners who lack prior experience with collaborative learning practices. Beyond scripts, research emphasizes the importance of shared artifacts as mediators of collaboration. Learning emerges through the joint creation and manipulation of representations such as code, diagrams, simulations, or narratives. Studies demonstrate that discourse and artifacts co-evolve, with learners' talk shaping artifacts and artifacts constraining and guiding subsequent interaction (Harrer et al, 2013; Mitnik et al, 2009). This artifact-centered view aligns with sociocultural perspectives and underscores the distributed nature of cognition in collaborative settings. At the classroom level, orchestration frameworks and integrated platforms support teachers in managing multiple collaborative groups simultaneously. These systems provide awareness of group progress, interaction patterns, and potential breakdowns, enabling timely intervention without excessive cognitive load (Harrer et al, 2013; Florido & Hernández-Leo, 2024; Zhao & Liu, 2024). Such work highlights the importance of balancing structure with learner autonomy, a recurring challenge in CSCL design.

As CSCL research matured, attention increasingly turned to learner and group characteristics. Empirical studies show that motivational constructs such as self-efficacy and collective efficacy strongly influence participation, discourse quality, and learning outcomes (Wang & Lin, 2007; Zhao & Liu, 2024). Groups with higher collective efficacy tend to engage in more sustained, higher-level interaction. These findings motivated extensive research on group formation and profiling. Approaches include constraint satisfaction models, optimization techniques, genetic algorithms, and machine learning methods that consider cognitive ability, personality traits, social relationships, and mentoring roles (Balmaceda et al, 2014; Costaguta, 2015; Olakanmi & Vassileva, 2017; Shawky & Badawi, 2018). Across these approaches, group formation is increasingly viewed as a pedagogical intervention rather than an administrative convenience. In parallel, advances in mobile and ubiquitous computing expanded CSCL beyond fixed classroom settings. Research on ambient intelligence and context-aware CSCL demonstrates how systems can adapt to learners' physical location, device availability, and situational constraints, enabling collaboration across time and space (García et al, 2012 & 2013; Uchida & Yuasa, 2019). Profile-enhanced learning frameworks further extend personalization by incorporating social, behavioral, and contextual dimensions into learner models (Chan et al, 2006).

The digitalization of collaborative learning environments enabled large-scale collection of interaction data, leading to the rise of learning analytics for CSCL. Early work focused on chat-based analytics to assess participation, estimate competence, summarize discussions, and identify pivotal moments in discourse (Dascalu et al, 2008 & 2010; Chiru & Trausan-Matu, 2012). These studies established that discourse features can serve as meaningful indicators of learning-related processes. Subsequent research incorporated machine learning and natural language processing to predict learning outcomes and diagnose collaboration quality from conversational cues and discourse sequences (Adamson et al, 2014; Carpenter et al, 2020). More recent work extends these approaches through multimodal learning analytics, integrating text, audio, video, physiological signals, and facial expressions to capture dimensions of collaboration that are otherwise difficult to observe (Nguyen et al, 2022; Dang et al, 2023; Ngo et al, 2024). However, empirical studies consistently highlight challenges related to noise, robustness, and ecological validity. Models trained on controlled datasets often degrade in authentic classroom environments, where sensor noise and contextual variability are common (Chejara et al, 2024; Garcia et al, 2013). These findings emphasize the need for analytics pipelines that are not only accurate but also deployable and resilient.

Building on analytics, a major research strand concerns adaptive collaborative learning support (ACLS), systems that monitor collaborative processes and intervene to improve interaction and learning. A unifying conceptualization frames these systems as operating through regulative loops, involving observation, comparison to pedagogical standards, and intervention through mirroring, assessment, or coaching (VanLehn 2016; Neto et al, 2022;). Empirical evidence suggests that ACLS can improve collaboration quality and learning outcomes compared to unsupported or non-adaptive collaboration (Suebnukarn & Haddawy, 2006; Haq et al, 2020). However, scholars caution against over-automation, warning that poorly designed adaptivity may undermine learner agency or impose rigid interaction patterns. Position papers contrast dystopian futures of opaque automation with utopian visions of theory-driven, transparent, and flexible systems (Rummel et al, 2016; Roschelle, 2021). Conversational agents increasingly serve as facilitators of collaboration, providing reflection prompts, discussion guidance, and orchestration support rather than direct instruction (Rosé & Ferschke, 2016; Naik et al, 2024; Gutiérrez-Ferré et al, 2024; Florido & Hernández-Leo, 2024). This trajectory aligns with recent work on generative AI, which positions AI as a co-orchestrator that augments human regulation and decision-making rather than replacing teachers or learners.

Across the CSCL literature, motivation and affect consistently emerge as mediating variables between design features and learning outcomes. Studies show that self-efficacy and collective efficacy influence participation, discourse quality, and persistence in collaborative settings (Wang & Lin, 2007; Zhao & Liu, 2024). Emotional alignment and regulation further shape how groups respond to challenges and sustain engagement. Gamification has been widely explored as a motivational strategy, yet results are mixed. Evidence suggests that gamification can enhance engagement and effort, but effects are inconsistent unless designs are personalized and grounded in motivational theory (Knutas et al, 2019; Agredo-Delgado et al, 2018; Tolmachova & Ilkou, 2022). These findings caution against one-size-fits-all gamification in collaborative environments. Embodied and game-based collaborative activities provide additional evidence that shared, engaging experiences can enhance conceptual understanding and collaboration quality (Mitnik et al, 2009; Zhao & Liu, 2024; Uchida & Yuasa, 2019). Collectively, this body of work supports a shift toward intelligence augmentation for collaboration, where AI systems enhance regulation, reflection, and coordination rather than optimizing isolated performance metrics (Roschelle, 2021).

## Related work and Theoretical Overview

Research on Socially Shared Regulation of Learning (SSRL) highlights how effective collaboration depends on groups' ability to align understanding, negotiate plans, and respond adaptively to challenges. Shared regulation involves cognitive, metacognitive, motivational, and emotional dimensions, such as sustaining collective efficacy, managing frustration, and maintaining engagement. In CSCL contexts, SSRL provides a key theoretical lens for designing scripts, tools, and adaptive supports that foster joint monitoring and reflection, enabling groups to take collective responsibility for their learning processes and outcomes. This perspective expands upon individual self-regulated learning by emphasizing the collective management of cognitive, motivational, and emotional resources within a group setting (Nguyen et al., 2025; Panadero et al., 2015). This shift acknowledges that while individual self-regulation is challenging, coordinating such processes within a group introduces even greater complexity and multifaceted challenges (Järvelä et al., 2019). Therefore, the mechanisms by which socially shared regulation of learning facilitates collaborative learning involve the joint employment of regulatory actions, strategic negotiation, and shared awareness among group members (Isohätälä et al., 2016; Silva et al., 2023). This involves the co-construction of goals, strategies, and reflections, where regulative acts and emerging perceptions are jointly evoked rather than individually directed (Ito & Umemoto, 2022). This collaborative dimension extends beyond mere individual contributions, integrating diverse regulatory behaviors like shared planning, monitoring, and evaluation, along with collective motivational and emotional regulation (Sulla et al., 2023; Yang et al., 2025).

The inherently interactive nature of regulation within collaborative learning contexts underscores the necessity of considering how individual intentions, beliefs, and prior sociohistorical experiences coalesce into a dynamic, collective regulatory process (Nguyen et al., 2025). Specifically, socially shared regulation of learning refers to the interdependent or collectively shared regulatory processes, beliefs, and knowledge that group members orchestrate to achieve a co-constructed or shared outcome (Guo et al., 2025). This involves group members jointly undertaking regulation activities for content processing and task completion (Moreno et al., 2016). This contrasts with co-regulation, which involves temporary coordination between individuals for regulating learning, or one individual guiding another, rather than a truly joint effort by multiple learners (Ito & Umemoto, 2022; Panadero et al., 2015). For instance, shared regulation encompasses instances where groups collectively define task perceptions, establish common goals, and engage in distributed regulatory actions that are negotiated until a consensus is reached (Järvelä & Hadwin, 2013).

The dynamic interplay allows groups to strategically adjust cognitive, motivational, and emotional processes, aligning their collaborative efforts with task goals and shared standards (Järvelä et al., 2019). This dynamic interplay allows groups to strategically adjust cognitive, motivational, and emotional processes, aligning their collaborative efforts with task goals and shared standards. This is particularly evident in the context of computer-supported collaborative learning, where groups must collectively regulate their cognition, motivation, emotions, and environment to ensure successful collaboration (Malmberg et al., 2015). This collective regulation is crucial for optimizing learning outcomes and fostering deep-level strategic engagement among collaborators (Järvelä & Hadwin, 2013; Panadero et al., 2015). Socially shared regulation of learning moves beyond individualistic approaches by acknowledging the pivotal role of motivation and emotions in collaborative learning, which are often overlooked in traditional cognitive-centric views (Malmberg et al., 2015). Consequently, it encompasses the intricate interplay of motivational, emotional, metacognitive, and strategic behaviors that are collectively orchestrated to ensure successful learning outcomes within a group (Järvelä et al., 2014). This framework helps to explain how collaborative learning mechanisms function by highlighting the critical role of social interaction in shaping and supporting learners' regulatory processes (Järvelä et al., 2014; Silva et al., 2023). What distinguishes socially shared regulation from co-regulation is the extent to which joint regulation emerges through a series of transactive exchanges among group members, emphasizing that students' interactions are key to enabling collective regulation (Sulla et al., 2023). This collective engagement is critical, as many learners struggle to develop the necessary regulatory skills independently, especially when confronted with complex collaborative tasks (Järvelä et al., 2016). SSRL, therefore, represents a multifaceted process where group members jointly construct strategies, such as planning, goal setting, monitoring, and evaluation, extending individual regulatory processes to a collective level (Yang et al., 2025).

## The Role of Eye-Tracking in Understanding Collaborative Processes in Computer-Supported Collaborative Learning

Eye-tracking technology offers an unparalleled opportunity to delve into the intricate cognitive and attentional mechanisms underpinning successful group interactions within these digital environments (Sharma et al., 2018). Specifically, dual eye-tracking allows researchers to analyze joint visual attention, which serves as a proxy for the quality of collaboration by indicating shared focus on common references and mutual monitoring of attention (Bryant et al., 2019). This method captures not only where individuals are looking but also how their gazes align over time, providing insights into the coordination of cognitive processes during shared tasks (Cheng et al., 2021). Furthermore, eye-tracking data can reveal how individuals re-evaluate their understanding based on differing perspectives encountered during collaboration, as indicated by shifts in gaze patterns towards abstract representations when reconciling divergent viewpoints (Steier & Davidsen, 2021). Such detailed analysis of gaze behavior facilitates a deeper understanding of socio-cognitive processes, such as grounding and mutual regulation, that are critical for effective collaborative knowledge construction (Sangin et al., n.d.). The synchronized recording of eye movements from multiple learners allows for the assessment of collaborative problem-solving processes and can guide improvements in the usability of CSCL environments (Uzunosmanoğlu & Çakır, 2014). This approach has gained traction as a valuable source of process data in educational research, covering diverse learning ecosystems from online to co-located and remote collaborative settings (Olsen et al., 2018). It also extends to understanding teachers' orchestration processes and providing adaptive feedback to students in real-time (Pedersen et al., 2018).

Eye-tracking’s non-invasive approach offers a unique window into cognitive processes during collaborative learning, overcoming the limitations of analyzing dialogue alone, which can be time-consuming and may undervalue moments of silence (Steier & Davidsen, 2021). Moreover, eye-tracking provides an automatic method for assessing collaboration, thereby enriching the understanding of collaborative cognition (Papadopoulos et al., 2017). While qualitative methods have traditionally dominated the study of collaborative cognition, eye-tracking data offers a quantitative alternative for assessing collaboration quality and success (Papadopoulos et al., 2017). Despite these advantages, merely observing that students look at the same object does not inherently confirm that they are engaging in joint problem-solving (Lämsä et al., 2022). Therefore, a complementary approach that augments joint visual attention with joint mental effort is necessary to provide deeper insights into the collaborative process (Steier & Davidsen, 2021). This dual eye-tracking methodology, which considers both visual attention and cognitive effort, offers a more comprehensive assessment of collaborative quality than either metric alone (Steier & Davidsen, 2021). This comprehensive perspective reveals that high-performing collaborative dyads often exhibit elevated levels of both joint visual attention and joint mental effort, indicating more equitable participation and potentially better task performance (Steier & Davidsen, 2021). Several studies have, for instance, demonstrated that aligned gazes between speaker-listener pairs correlate with listener comprehension, and productive collaboration is associated with high joint visual recurrence (Schneider et al., 2013).

Such results suggest that eye-tracking can be an effective tool for identifying instances of effective communication and shared understanding within collaborative learning contexts (Uzunosmanoğlu & Çakır, 2014). Additionally, eye-tracking can elucidate the impact of affective states, such as boredom or curiosity, on collaborative dynamics, thereby informing timely interventions within adaptive learning systems (Olsen et al., 2018). Moreover, integrating gaze awareness tools within CSCL environments has been shown to foster joint attention and enhance the quality of collaboration by providing real-time feedback on partner's gaze behaviors (Hayashi, 2020). These advancements enable a more nuanced analysis of social interactions, moving beyond simple self-reported data to capture dynamic processes like joint visual attention, which predicts successful collaboration (Barmaki & Guo, 2020). For example, studies have shown that pairs of students demonstrating higher levels of sustained joint attention, particularly within intelligent tutoring systems, often achieve greater learning gains (Reuscher et al., 2023). This further suggests that joint visual attention serves as a reliable indicator for the quality of collaboration, providing an objective measure for assessing effective interaction in educational settings (Papadopoulos et al., 2017). This understanding extends to how the even distribution of initiating joint visual attention moments among group members correlates with enhanced learning outcomes (Papadopoulos et al., 2017). Beyond visual coupling, mutual gaze, and gaze aversion, the concept of joint mental effort, derived from pupil-based data, offers another lens through which to examine collaborative dynamics by quantifying the similarity in cognitive load between team members (Reuscher et al., 2023). This integrated approach allows researchers to explore the causal relationships between joint mental effort and joint visual attention, revealing how these factors contribute to effective co-construction and knowledge building in collaborative tasks (Sharma & Olsen, n.d.). Specifically, integrating gaze indicators into collaborative interfaces can enhance communication by making partners aware of each other's interests and attention, although a balance between visibility and potential distraction must be maintained (Cheng et al., 2021). For example, a knowledge awareness tool has demonstrated that learners episodically refer to such cues, primarily to evaluate the epistemic value of information provided by a peer, particularly when uncertainty is perceived (Sangin et al., n.d.). Further, research indicates that providing visible gaze feedback to learners, especially when combined with metacognitive suggestions from a pedagogical conversational agent, significantly improves coordination and learning performance (Hayashi, 2020). This is particularly evident in studies where learners provided with their partner's gaze information achieved higher quality collaboration and improved learning outcomes compared to control groups (Schneider et al., 2013). Despite these advancements there is still an need to understand the optimal modalities for presenting gaze awareness to avoid cognitive overload while maximizing its utility in fostering joint attention and mitigating misunderstandings during remote collaborative problem-solving (Cheng et al., 2021; Hayashi, 2020). This challenge is amplified in complex collaborative scenarios where multiple participants engage in nuanced interactions, necessitating sophisticated aggregation and visualization techniques to prevent information overload (Cheng et al., 2021).

## Eye Tracking and Mechanisms Underlying Pair Programming

Eye-tracking technology offers a nuanced lens into the cognitive and collaborative processes that underpin pair programming, allowing researchers to observe programmer attention and interaction patterns at a granular level (Bansal et al., 2023). Specifically, dual eye-tracking setups enable the simultaneous capture of visual attention from both participants, providing insights into joint attention, gaze coupling, and shared understanding during collaborative coding tasks (Pietinen et al., 2008; Villamor & Rodrigo, 2018). This methodology is particularly valuable for discerning the intricate communication dynamics and individual expertise within pair programming teams, ranging from novices to experts (Jang et al., 2024; Villamor & Rodrigo, 2018). For instance, researchers have employed eye-tracking to identify how communication skills evolve across different group compositions, noting a tendency for participants to prioritize code exploration, especially during challenging tasks (Jang et al., 2024). This allows for a deeper understanding of how visual cues, alongside verbal communication, facilitate problem-solving and knowledge transfer within a pair programming context (Pietinen et al., 2010; Villamor & Rodrigo, 2018). Furthermore, eye-tracking can delineate how visual attention shifts between different code elements and documentation, reflecting varying cognitive loads and comprehension strategies between expert and novice programmers (Aljehane et al., 2023; Bieliková et al., 2018). This precise tracking of visual focus can reveal how developers allocate cognitive resources to different parts of the code, indicating areas of confusion or concentration (Tang et al., 2024). Moreover, eye-tracking data can reveal how shared visual attention impacts collaboration, identifying instances where synchronized gaze patterns correlate with effective problem-solving and knowledge exchange within programming dyads (Pietinen et al., 2010). This enables the analysis of collaboration effectiveness through metrics derived from cross-recurrence plots and their associated quantification analysis, which can distinguish between successful and unsuccessful programming pairs based on their gaze collaboration patterns (Villamor & Rodrigo, 2019). Dual eye-tracking studies have consistently highlighted the importance of synchronized visual attention in unveiling the cognitive mechanisms underlying successful collaboration, with high cross-recurrence in gaze patterns correlating with enhanced comprehension and task success in various collaborative scenarios (Magnussen et al., 2017; Villamor & Rodrigo, 2019). This granular analysis of gaze patterns, particularly through concepts like gaze coupling, can differentiate between convergent phases where partners jointly understand code and divergent phases where individual comprehension is developed (Villamor & Rodrigo, 2022). Such fine-grained analysis is crucial for understanding how expertise levels influence visual attention strategies, with novices often exhibiting more fixations and longer gaze durations on code compared to experts, reflecting higher cognitive effort (Aljehane et al., 2023). In addition, analyzing visual attention measures such as fixations and saccades, along with the time spent on specific Areas of Interest, can provide insights into complex cognitive activities (Papavlasopoulou et al., 2018). These measures can quantify cognitive load, problem-solving strategies, and the allocation of attention during code comprehension and debugging tasks, thus offering a window into the nuanced mental processes of programmers (Jang et al., 2024). Furthermore, this methodology can differentiate between how experts and novices approach code comprehension tasks, revealing that experts tend to read code more efficiently and with lower cognitive load (Aljehane et al., 2023; Peitek et al., 2022). Moreover, eye-tracking has been utilized to detect misconceptions during collaborative programming by observing instances where a developer's gaze deviates from the intended focus, despite verbal or gestural cues, particularly when multiple identical elements exist in the code (Pietinen et al., 2010). This capacity to detect divergent attention in such specific scenarios underscores the utility of eye-tracking in identifying subtle breakdowns in shared understanding that might otherwise go unnoticed in traditional observational studies. These methods also extend to predicting task success by analyzing gaze density and fixation dispersion, further enhancing our understanding of collaborative efficacy (Magnussen & Elming, 2017). Such detailed insights into visual processing and shared attention are invaluable for testing hypotheses regarding a software developer's comprehension processes and cognitive workload (Aljehane et al., 2023). This is consistent with the "eye-mind hypothesis," which posits a direct link between visual attention and cognitive processing, allowing for inferences about cognitive states from eye movements (Villamor & Rodrigo, 2018). Therefore, eye tracking serves as a powerful tool for investigating cognitive processes, offering objective data to complement subjective self-reports and providing a deeper understanding of how programmers interact with code and with each other (Jessup et al., 2021; Peitek, 2018).

## Research Questions and Contributions

Grounded in research on socially shared regulation of learning (SSRL), collaborative learning analytics, and AI-driven adaptive support in CSCL, this work investigates how moment-to-moment coordination of attention and cognitive effort relates to collaborative performance and how AI-driven feedback can foster more productive regulation processes. Specifically, we examine pair programming as a complex collaborative task requiring continuous alignment of cognitive, motivational, and attentional processes. Using joint visual attention (JVA) and joint mental effort (JME) as process-level indicators of shared regulation, we address the following research questions:

**RQ1**. How are joint visual attention (JVA) and joint mental effort (JME) associated with collaborative performance in pair programming, and how do these processes differ between high- and low-performing dyads?

**RQ2**. What is the causal relationship between JVA and JME during collaborative problem solving, and how does this causal structure differ between successful and unsuccessful collaboration?

**RQ3.** How does a reactive, real-time feedback based on JVA and/or JME influence collaborative performance, shared regulatory processes, and the distribution of JVA–JME interaction episodes?

**RQ4.** To what extent can a proactive, forecast-based adaptive feedback improve collaborative processes and performance compared to both no feedback and reactive feedback alone?

Together, these research questions aim to clarify both the mechanisms underlying effective socially shared regulation and the design principles for adaptive, AI-supported collaboration in CSCL environments.

This paper makes the following contributions to CSCL research and the study of socially shared regulation of learning:

**Empirical characterization of shared regulatory processes in pair programming**: We provide fine-grained empirical evidence that both joint visual attention and joint mental effort are strongly associated with collaborative performance. High-performing dyads exhibit significantly higher JVA and JME, as well as a greater prevalence of productive high-JME–high-JVA episodes, offering process-level validation of SSRL theory in an authentic programming task context

**Causal evidence linking cognitive alignment to attentional coordination:** Through Granger causality analysis, we demonstrate that in successful collaboration, joint mental effort systematically precedes and predicts joint visual attention. This finding advances prior correlational work by showing that shared cognitive engagement is a driving mechanism for attentional alignment, rather than merely a by-product of it, thus refining theoretical accounts of regulation dynamics in CSCL.

**Design and evaluation of multimodal, regulation-aware adaptive feedback:** We introduce and empirically test a suite of adaptive feedback mechanisms, including gaze awareness, dialogue prompts, task-based hints, and AI code assistance, triggered by deviations in JVA and JME. Results show that combined feedback targeting both attention and mental effort produces the strongest gains in collaboration quality and performance, compared to single-channel or no-feedback conditions.

**Demonstration of proactive intelligence augmentation for collaboration:** By integrating machine-learning-based forecasting of collaborative states, we show that proactive feedback can outperform reactive support by anticipating breakdowns in shared regulation before they fully manifest. This contribution extends ACLS research by illustrating how predictive analytics can support timely, minimally disruptive interventions that preserve learner agency.

**Methodological advancement for studying SSRL in real time:** The paper introduces a robust methodological framework that combines dual eye-tracking, pupillometry, episode-based analysis, and causal modeling to study SSRL as a dynamic, emergent process. This approach addresses known challenges in ecological validity and provides a scalable pathway for deploying learning analytics-driven regulation support in authentic CSCL settings.

Collectively, these contributions position AI not as an automated controller of collaboration, but as an intelligence-augmenting partner that supports learners' capacity to regulate attention, effort, and understanding together.

# Methodology

In this section, we will present the three studies included in this paper. However, first we will present the core concepts that are common to two or more studies before presenting the individual studies. First, we will present the performance and eye-tracking measurements and episodes using them (used in all the three studies). Second, we will present the different feedback/scaffolding tools designed (used on the second and the third studies). Then we will present the analysis methods (common across all the studies). Finally, we will present the three studies.

## Performance measure: debugging success

The total number of bugs successfully fixed in the given task time. This metric measures the effectiveness of the pair in correctly identifying and resolving errors. The score is increased by one, indicating the number of bugs solved plus the bug the participants were working on as the task ended.

## Eye-tracking measurements and episodes

### Joint visual attention

JVA is the amount of time the peers spend looking at the same set of objects in a given time window. The quantification of JVA is achieved by comparing the aligned attention patterns of two collaborators on a shared code document. This method overcomes issues from screen dynamics (like scrolling or terminal resizing) by establishing a persistent, non-visible grid based on code structure, vertically by six-codeline intervals and horizontally by screen percentiles. Raw eye-gaze coordinates are first mapped to their corresponding codeline by converting the gaze percentage to a pixel coordinate, which is then correlated with the known display position of the code document. The frequency of gaze occurrences in each grid slot is accumulated over a short interval (30 seconds) for both participants. Finally, the JVA score, which represents the degree of their shared attentional focus, is calculated using cosine similarity between the two resulting gaze frequency distribution grids.

### Joint mental effort

Each participant's cognitive load was quantified using the Index of Pupillary Activity (IPA) Duchowski et al 2020, a wavelet-based signal decomposition of the pupil diameter time series that captures high-frequency oscillations linked to mental-effort. The resulting ME signals (ME1 and ME2) were segmented into non-overlapping 10-second windows. Within each window, JME was computed using cosine similarity, producing a time-aligned similarity score that reflects how closely the two partners' cognitive load patterns evolved over the same interval. Higher values indicate stronger synchrony in mental-effort, whereas lower values reflect divergence or imbalance in the pair's engagement. JME, defined as cognitive similarity or the similarity in mental-effort between the two peers in the dyad, is calculated server-side from the time series of the individual ME (ME1 and ME2) values to compute JME, the resulting individual ME time series are first discretized into a manageable integer range. The similarity, or synchrony, between these two discretized time-series is then calculated using cross-recurrence quantification (Coco et al 2014), which yields the JME score that represents the degree to which the dyad is sharing a similar cognitive state across time.

### JVA-JME episodes

We computed the JME-JVA episodes based on whether the JME was high or low in a given time window of ten seconds, and whether the JVA was high or low in the same time window of ten seconds (Sharma et al, 2018). where this duration was deemed suitable for defining such episodes. The high or low JME was determined by using a median-cut for the JME for each dyad. Similarly, the high or low JVA was determined by using a median-cut for the JVA for each dyad. Finally, there were four types of episodes: high JME low JVA, high JME high JVA, low JME high JVA, and low JME low JVA.

## Feedback tools

### Github Co-pilot

In this study we employ only GitHub Copilot's AI-driven autocomplete, a feature that generates context-aware code suggestions directly within the editor. Prior work shows that Copilot can accelerate task completion and reduce perceived cognitive effort [38] , though its influence on learning remains underexplored. Within our framework, "enabling Copilot" refers specifically to activating this autocomplete mechanism as a momentary scaffold when forecasted mental-effort patterns indicate cognitive strain. The demonstration of this feedback can be seen in Figure 1.

```
100
101         # Do we bounce off the left of the screen?
102         if self.x <= 0:
103             self.direction = (360 - self.direction) % 360
104             self.x = 1
105
106         # Do we bounce of the right side of the screen?
107         if self.x > self.screenwidth - self.width:
108             self.direction = (360 - self.direction) % 360
109             self.x = self.screenwidth - self.width - 1
110
111         # Did we fall off the bottom edge of the screen?
112         if self.y > 600:
                return True
```

*Figure 1: Github auto-complete suggestion (the last two lines of the code in this image).*

## Dual text selection

Broadcasted text selection is implemented as an always-on mechanism in which each collaborator's cursor selection is continuously shared, offering a lightweight visual cue that facilitates rapid joint focus. Modern shared editors, including Visual Studio Code's Live Share used in our experiment, inherently support dual text selection, enabling both partners to highlight code independently while preserving mutual awareness. This design aligns with the role of Joint Visual Attention (JVA), the similarity of partners' gaze patterns, as a core indicator of coordination in concurrent collaboration. Figure 2 illustrates this feedback visualization.

```
100
101         # Do we bounce off the left of the screen?
102         if self.x <= 0:
103             self.direction = (360 - self.direction) % 360
104             self.x = 1
105
106         # Do we bounce of the right side of the screen?
107         if self.x > self.screenwidth - self.width:
108             self.direction = (360 - self.direction) % 360
109             self.x = self.screenwidth - self.width - 1
110
111         # Did we fall off the bottom edge of the screen?
112         if self.y > 500:
113             self.direction = (180 - self.direction) % 360
114
```

*Figure 2: Dual text selection, in this image, we see blue being the self-selection and purple being the partner's selection.*

## Gaze-awareness tool

A translucent, colored rectangle indicates the partner's current gaze region (approximately nine code lines).activating visual feedback only when the relevant synchrony metric falls below a predefined threshold. The visualization of this scaffolding can be found in Figure 3. This feedback is activated when the JVA is low.

*Figure 3: Gaze awareness feedback (the green bar on the left side of the code)*

## Dialogue prompt

This feedback is intentionally designed as the least disruptive form of support and is triggered when the pair's JME is low, indicating a possible asymmetry in understanding. When activated, a small unobtrusive prompt appears in the bottom-right corner of the editor (Figure 4), encouraging the partners to initiate brief dialogue. This conversational nudge is intended to strengthen shared understanding and guide the pair toward more aligned and more optimal mental effort levels.

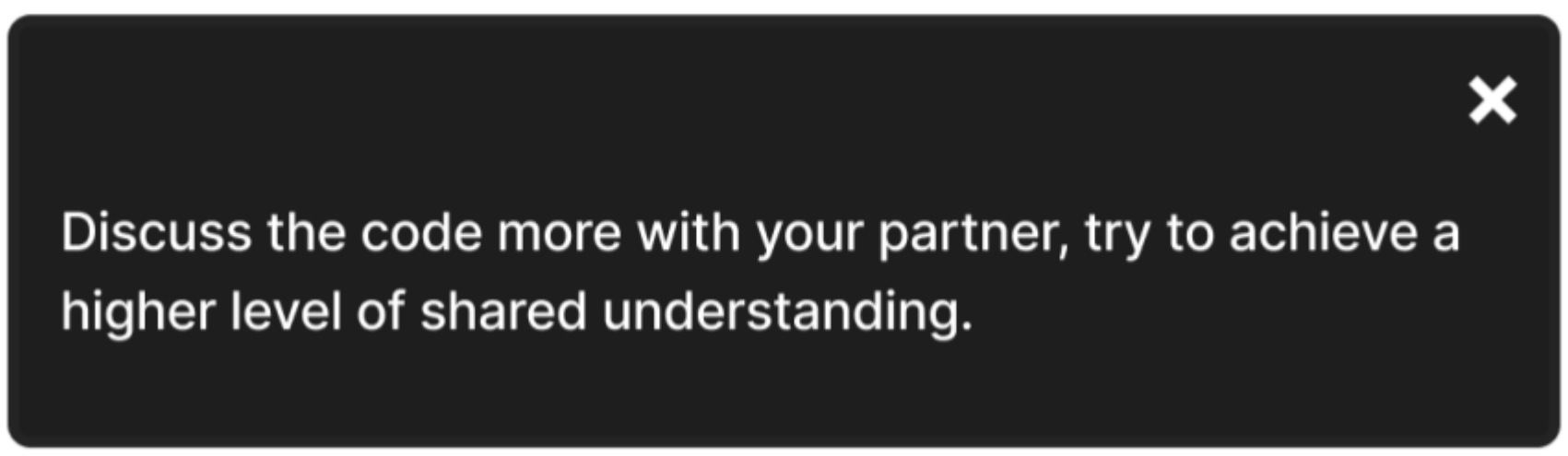

*Figure 4: The dialogue prompt shown to the participants.*

### Task-based hint

The task-based hint constitutes the most disruptive form of support and is therefore reserved for situations where None of the previous scaffolding have been productive and both collaborators exhibit extreme Mental Effort levels, indicating substantial cognitive strain or stagnation. When triggered, the system presents a small window in which the user selects the current task and the specific bug, after which a targeted hint is provided (Figure 5). By directing the pair toward a relevant section of the code and clarifying the underlying issue, the hint is designed to realign their problem-solving trajectory and restore cognitive balance. Although more intrusive than other feedback types, this intervention is intended to produce a strong corrective effect, particularly during episodes of prolonged difficulty.

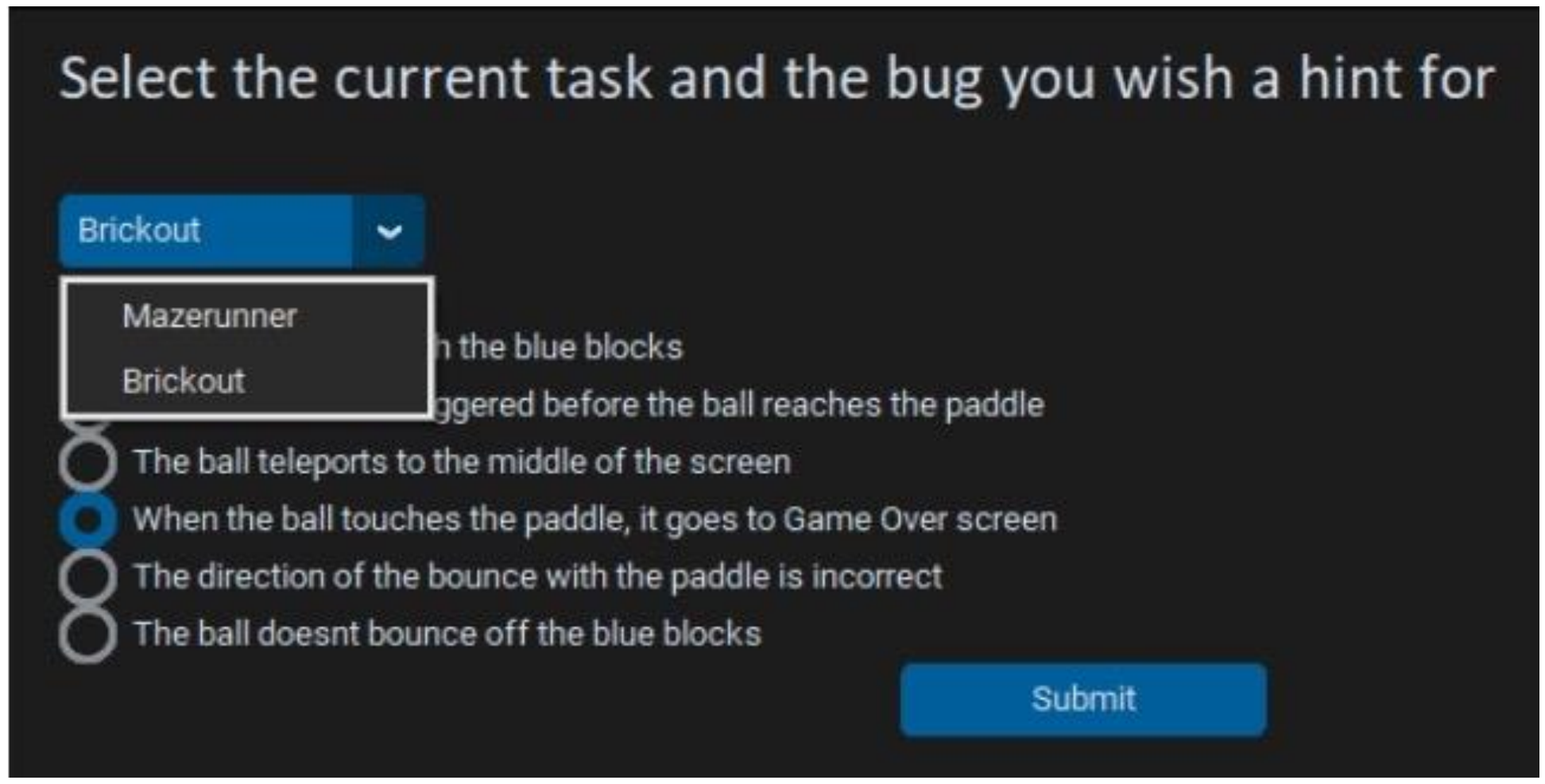

*Figure 5: task-based hints provided to the participants.*

## Causal modelling between JVA and JME

In this contribution, we analyse the causal link between JVA and JME. The methodology is described in full detail in Sharma et al (2021a) and (2021b). In the following, we briefly describe the methods to analyse causal link between two time series from multiple participants. To identify causal relations between JVA and JME, we used the Granger causality test. Granger causality (Granger, 1969) has two assumptions, first is that cause occurs before effect and second is that the cause has information about the effect that is more important than the history of the effect. Granger causality was selected since it has proven usefulness in the context of learning technologies (Sharma et al, 2021a and 2021b). For more details and the mathematical formulation of Granger Causality in the context of learning technologies please see Sharma et al, 2021b. Here we provide a summary of steps so the method can be replicated based on this paper only. Let'sassume that we are modelling the Granger causality between two variables X and Y. For each group compute the partial $\eta^2$ for the model "X Granger causes Y", the partial $\eta^2$ for the model "Y Granger causes X", the partial $\eta^2$ for the model "Y linearly predicts X" (this is similar to a correlational model), the difference between the $\eta^2$ of the two Granger causal models (this is the effect size or the strength of the Granger causality where a positive value indicates "X Granger causes Y" while a negative value indicates "Y Granger causes X") and the difference between the Granger causal model with higher $\eta^2$ and the $\eta^2$ of the correlational model (this is the significance of the Granger causality). Once we have the effect size and the significance of the Granger causality for each group, plot them on a Cartesian-coordinate system (Figure 6).

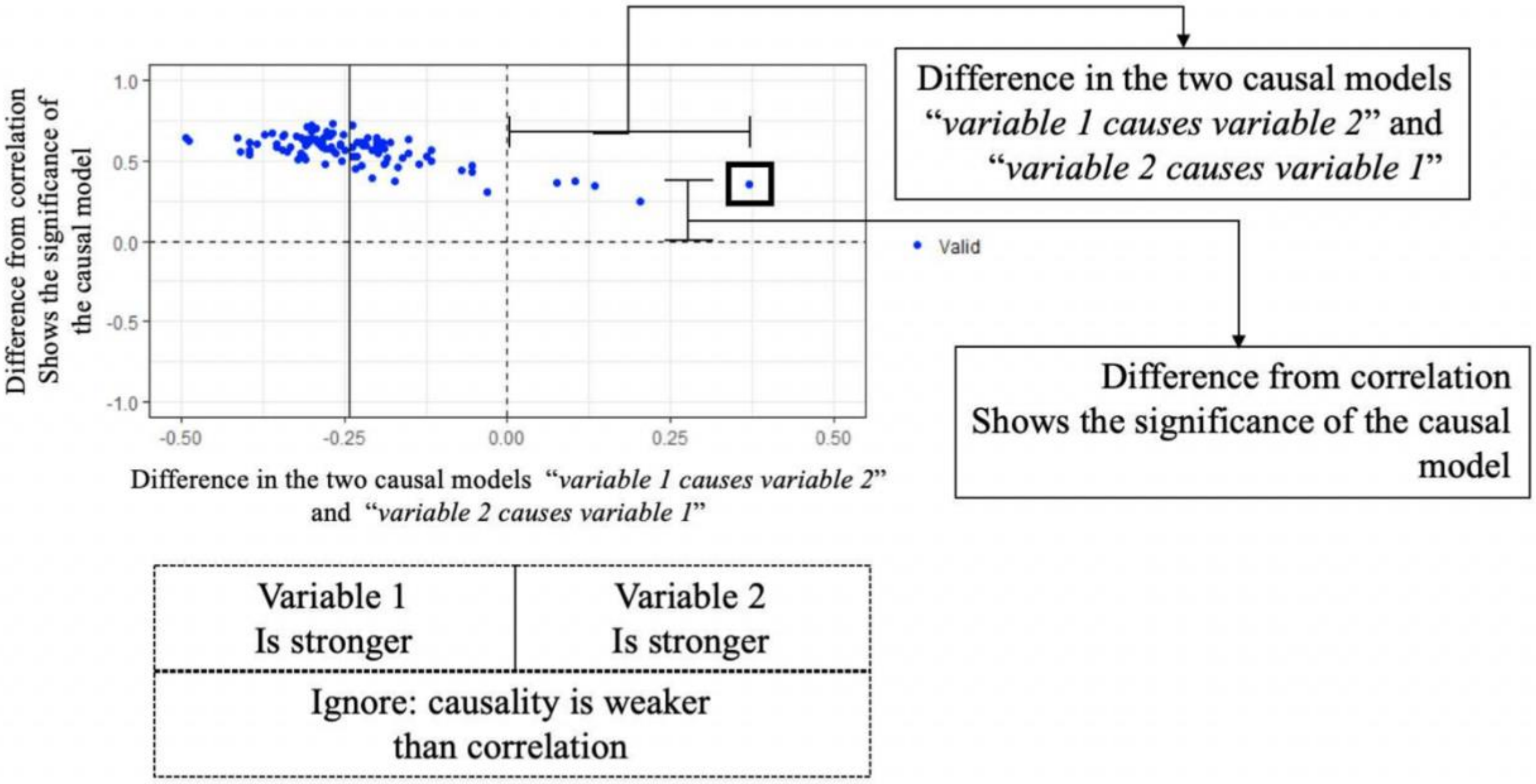

*Figure 6: A visualization to summarize causality results for multiple dyads. For dyads, we calculate the difference between the two causal models ('x causes y' and 'y causes x') and the respective difference between causal and correlational models.*

## Experiments

In this section, we present the details about the three studies. It is important to point out here that the tasks in the three studies had similar difficulty levels.

### Study 1: Basic pair programming

In this study, we employed a simple exploratory design without any experimental condition. We recruited 36 pairs (72 participants, 32 females, mean age 23.6 years, variance age 2.4 years) to participate in the pair programming tasks. The pairs were randomly formed. The participants were welcomed into the laboratory as pairs, where they first signed a consent form. Then the two eye-tackers were calibrated for the individual programmer. Finally, they were shown the correct version of the program that they received to debug and were provided the bug-infused version of the code to solve. The participants were informed that there were only logical bugs in the code and no syntax errors. In this study, we computed the debugging success and then applied a median-split to divide the pairs into high-performing and low-performing pairs.

### Study 2: Reactive feedback

In this study, we employed a between-subject design with four conditions: 1) control (no feedback); 2) JVA feedback; 3) JME feedback; 4) Both (combined JVA and JME) feedback. In each of the conditions, we recruited 30 pairs (120 pairs in total, 240 participants, 81 females, mean age 21.4, variance age 3.2) to participate in the pair programming tasks. The pairs were randomly formed. The participants were welcomed into the laboratory as pairs, where they first signed a consent form and then received a demonstration of the feedback tool they would be using. In the demo, they were informed about the interactions with the feedback tools, including when they would be triggered, what changes would be made in the code editor, and how they could switch the feedback on and off. Then the two eye-tackers were calibrated for the individual programmer. Finally, they were shown the correct version of the program that they received to debug and were provided the bug-infused version of the code to solve. The participants were informed that there were only logical bugs in the code and no syntax errors.

The reactive component responds to the immediate, moment-to-moment conditions of the pair. Feedback is triggered whenever the real-time measurements deviate more than 2 standard deviations (2SD) from each participant's resting baseline, enabling rapid correction when the system detects abrupt or pronounced shifts. The detailed scenario logic and feedback combinations are shown in Appendix A. For the combined JVA and JME condition (**both**), we used three feedback tools: offering help by enabling the gaze-awareness tool, prompting dialogue initiation, and providing a task-based hint. For the JVA-only condition, we use two feedback tools: offering help through enabling the gaze-awareness tool and a prompt to initiate dialogue. For the JME-only condition, we used only one feedback tool: providing a task-based hint. The triggers are described in Appendix A.

### Study 3: Proactive feedback

In this study, we employed a within-subject design with two conditions: 1) control (no feedback) condition;  2) experimental (combined JVA and JME forecasting feedback) condition. We recruited 26 pairs to participate in the pair programming tasks. The pairs were randomly formed. The participants were welcomed into the laboratory as pairs, where they first signed a consent form and then received a demonstration of the feedback tool they would be using. In the demo, they were informed about the interactions with the feedback tools, including when they would be triggered, what changes would be made in the code editor, and how they could switch the feedback on and off. Then the two eye-tackers were calibrated for the individual programmer. Finally, they were shown the correct version of the program that they received to debug and were provided the bug-infused version of the code to solve. The participants were informed that there were only logical bugs in the code and no syntax errors.

The proactive component anticipates future collaboration states by relying on forecasted values of the system's key measures. An Extreme Gradient Boosting (XGBoost) model generates real-time predictions over a 30-seconds horizon, allowing the system to intervene before suboptimal states fully materialize. The feedback logic operated on collaboration states forecasted 30 seconds in advance by an XGBoost model. The model's predictions for JVA, JME, and ME were discretized into High (H), Average (AVG), and Low (L) categories and then compared against a desired collaboration matrix (see Table A1) in Appendix A). 2 The detailed scenario logic and feedback combinations are shown in Appendix A. For the feedback condition, we employed the complete logic presented in the Appendix.

## Analysis

To answer the first part of RQ1 (collaborative performance vs JVA and JME), we conducted ANOVA tests using the debugging performance levels, from the study 1, as independent variable and dual eye-tracking measures (JVA and JME) as the dependent variables. To answer the second part of RQ1 (collaborative performance vs JME-JVA processes), we conducted ANOVA tests using the debugging performance levels as independent variable and the proportions of JVA JME episodes (high JME high JVA, high JME low JVA, low JME high JVA, low JME low JVA) as the dependent variables.

To answer RQ2, we used the method based on Granger Causality presented above between JVA and JME. We present the causal analysis separately for the high and low performing pairs from study 1.

To answer RQ3, we used the data from study 2 (reactive feedback). We used ANOVA tests with the feedback conditions (control, JVA-only, JME-only, both JVA and JME), as independent variable. For different tests, we use JVA, JME and proportions of JME-JVA episodes as the dependent variables. Further pairwise ANOVA, for each ANOVA test, was used to find where the significant differences are originating from. We also analyse the causal relation between JVA and JME for the four feedback conditions as well to understand the interplay between the shared regulatory processes.

To answer the first part of RQ4 (effect of proactive feedback), we used the data from study 3 (proactive feedback). We used ANOVA tests using the feedback conditions (control vs experimental) as independent variable and JVA, JME and proportions of JME-JVA episodes as the dependent variables. We also analyse the causal relation between JVA and JME for the control and experimental conditions as well to understand the interplay between the shared regulatory processes.

To answer the second part of RQ4 (comparing the studies), we used three subsets: 1) , 2) , and 3) The experimental feedback data from the third study. We compare using ANOVA with Welch correction (due to different number of dyads in the three datasets). We use the datasets as the independent variable and JVA/JME, causal relations between JVA and JME, and proportions of JME-JVA episodes as dependent variables. Further pairwise ANOVA was used with Welch correction to find where the significant differences are originating from.

Whenever an ANOVA was use, prior to this, we used a Levens's test to verify homoscedasticity on the dependent variables and Shapiro-Wilk test to verify the normality. For the multiple comparisons, the p-value was adjusted using Bonferroni corrections.

# Results

## Study 1: Basic pair programming

We analysed the relationship between JVA and JME and the performance level of our dyads. For both measures, we observed significant associations with performance level. The JVA for high performing dyads is significantly higher than the JVA for the low performing dyads ($F[1,34] = 21.69$, $p < .0001$, Figure 7 left panel). Similarly, the JME for high performing dyads is significantly higher than the JME for the low performing dyads ($F[1,34] = 39.14$, $p < .0001$, Figure 7 right panel).

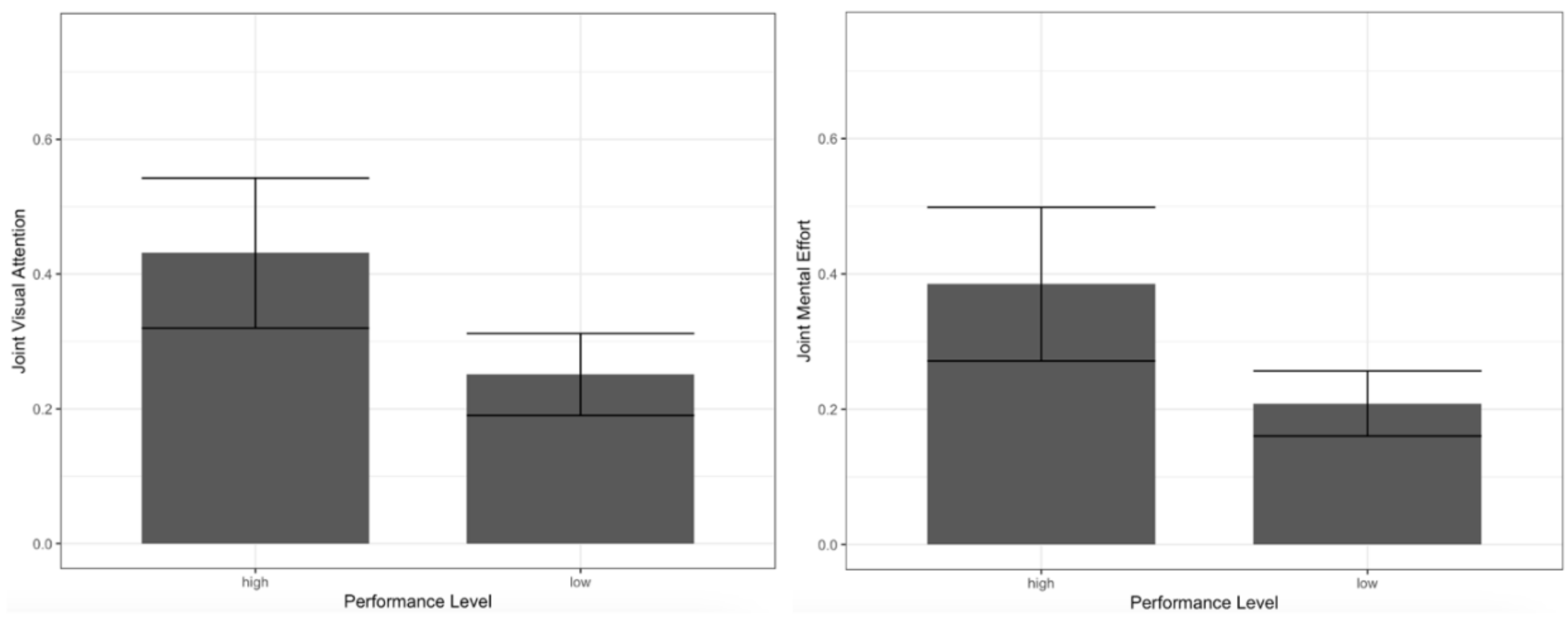

*Figure 7: JVA and JME for the two performance levels in the study 1*.

Next, we analyse the causal link between the JME and JVA for all the dyads in the first study. We observe from figure 8 that all the high performing pairs are in the first quadrant, that is, for all the high performing pairs "JME causes JVA" and this causality explains the relation between JME and JVA better than the correlation between the two measurements. We also observe that all the low performing pairs are either in the second or the third quadrant, that is, we can not claim any causal relation between JME and JVA because none of the low performing pairs the correlation between JME and JVA explains the relation between these better than the causal model.

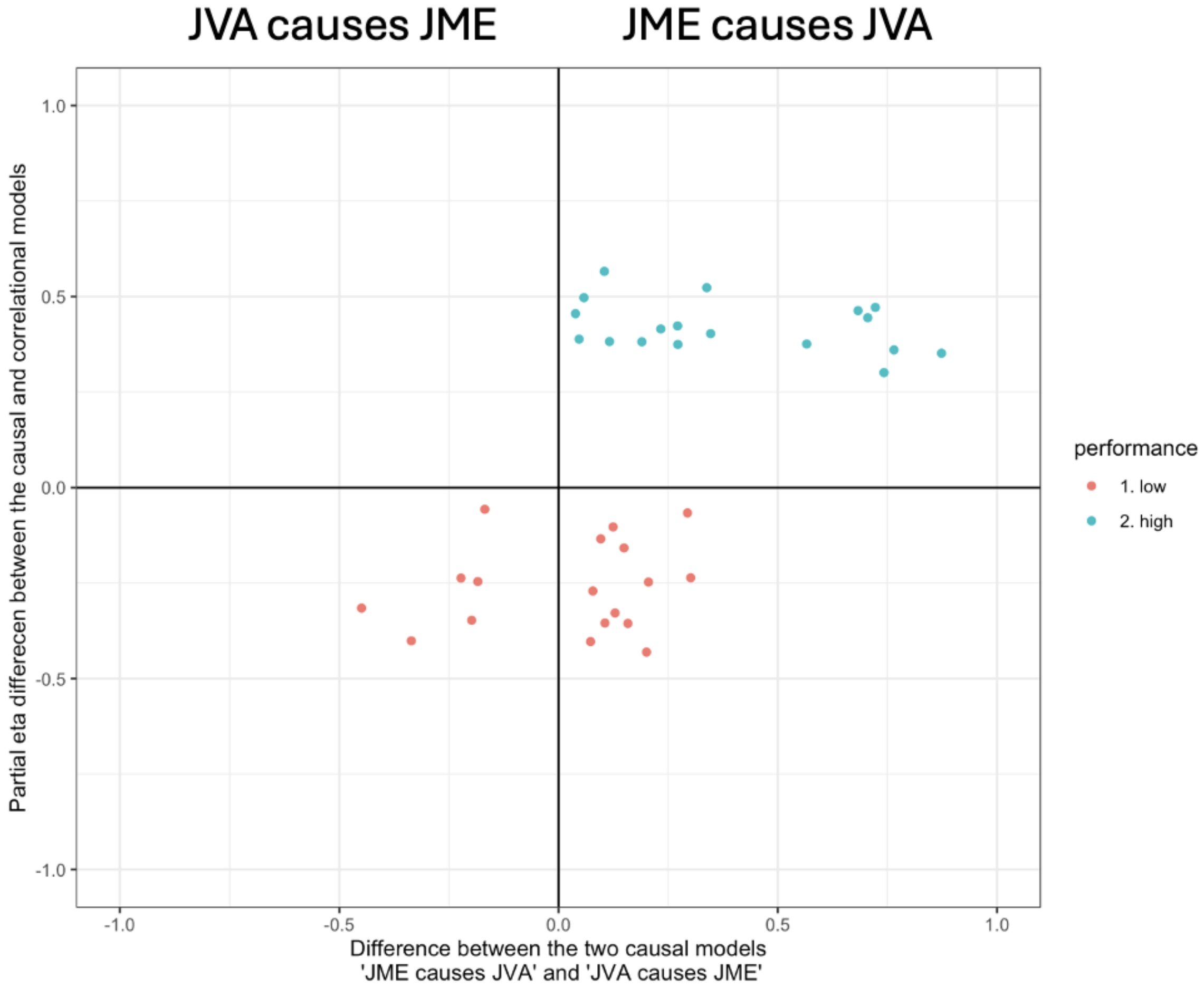

*Figure 8: causal analysis for the JVA and JME for the two performance levels in the study 1.*

Finally, we analyse the lengths of the JVA-JME episodes for high and low performing dyads (Figure 9). We observe that the high performers have higher proportion of “high JME high JVA” episodes than the low performers ($F[1,34] = 226.52$, $p < .0001$); while the low performers have higher proportions of “high JME low JVA” ($F[1,34] = 23.11$, $p < .0001$), “low JME high JVA”($F[1,34] = 36.01$, $p < .0001$), and “low JME low JVA” ($F[1,34] = 152.29$, $p < .0001$) episodes.

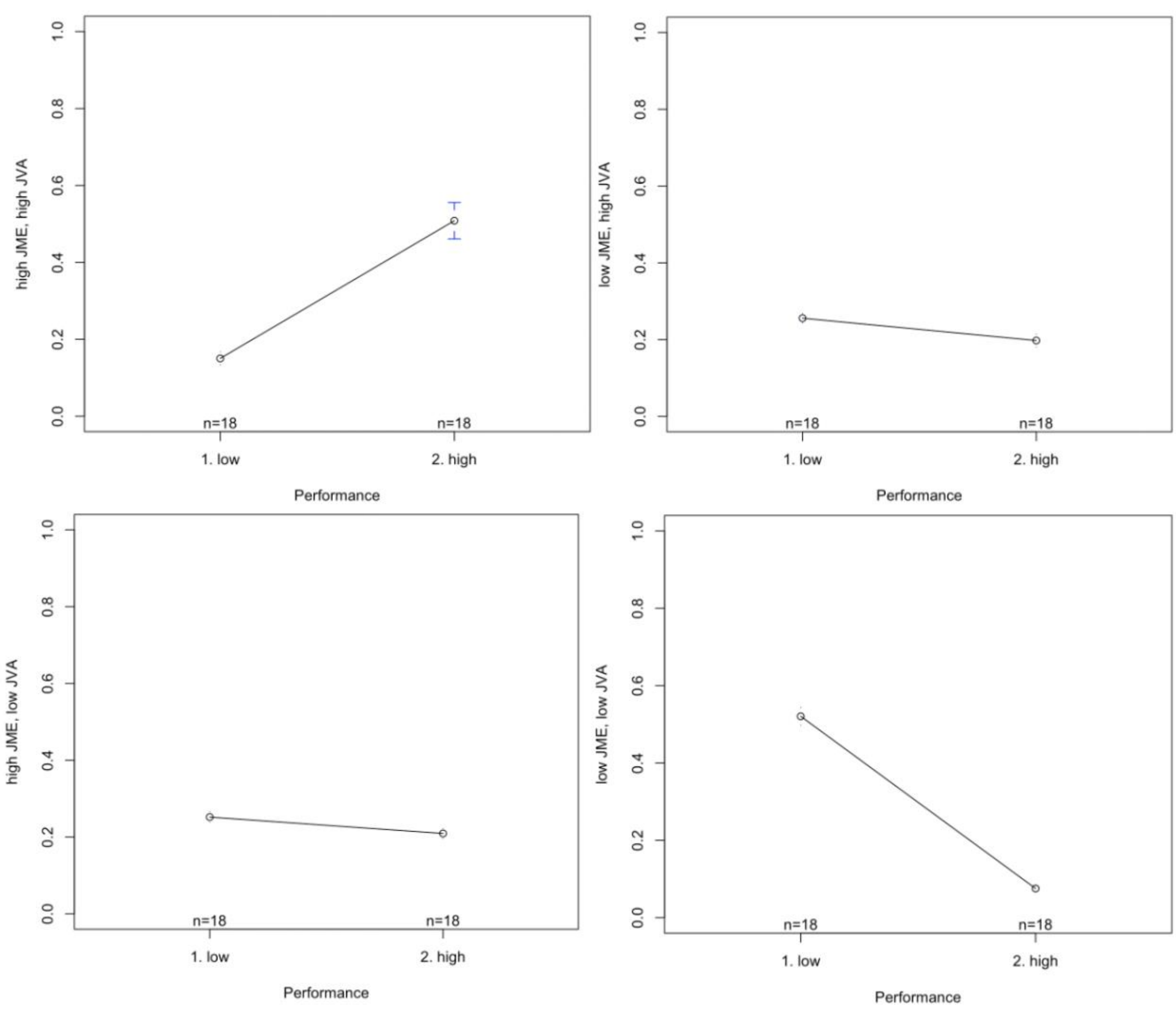

*Figure 9: proportions of the different JME-JVA episodes for the two performance levels in the study 1.*

## Study 2: Reactive feedback

For the reactive feedback tool, there is a positive impact on the debugging success (F[3,116] = 214.87, p < .0001). All the three feedback conditions improve the debugging success significantly than the control condition (Table 1, columns 3 and 4). Furthermore, both condition had the highest improvement followed by JME only condition (Table 1, columns 3 and 4) and finally the JVA only condition.

*Table 1: pairwise ANOVA comparisons for performance, JVA and JME for the four reactive feedback conditions (study 2).*

| | | **Performance** | | **JVA** | | **JME** | |
|---|---|---|---|---|---|---|---|
| | | **F** | **P** | **F** | **P** | **F** | **p** |

| **Control** | **JVA only** | 66.58 | <.0001 | 177.18 | <.0001 | 139.07 | <.0001 |
|---|---|---|---|---|---|---|---|
| **Control** | **JME only** | 236.2 | <.0001 | 104.32 | <.0001 | 308.81 | <.0001 |
| **Control** | **Both** | 598.91 | <.0001 | 332.68 | <.0001 | 561.5 | <.0001 |
| **JVA only** | **JME only** | 50.97 | <.0001 | 12.63 | .0007 | 67.92 | <.0001 |
| **JVA only** | **Both** | 261 | <.0001 | 37.43 | <.0001 | 219.48 | <.0001 |
| **JME only** | **Both** | 81.29 | <.0001 | 90.76 | <.0001 | 26.67 | <.0001 |

Next, we analyse the relation between JVA and the reactive feedback conditions (Figure 10). We observe that there is a significant relationship between the feedback condition and JVA ($F[3,116] = 141.32$, $p < .0001$). Further, the pairwise comparisons (Table 1, columns 5 and 6) show that the control condition has the lowest JVA, followed by JME only, JVA only feedback conditions, and finally "both" condition has the highest JVA. Similarly, we observe a significant relationship between the feedback condition and JME ($F[3,116] = 222.56$, $p < .0001$). Further, the pairwise comparisons (Table 1, columns 7 and 8) show that the control condition has the lowest JVA, followed by JVA only, JME only feedback conditions, and finally "both" condition has the highest JME.

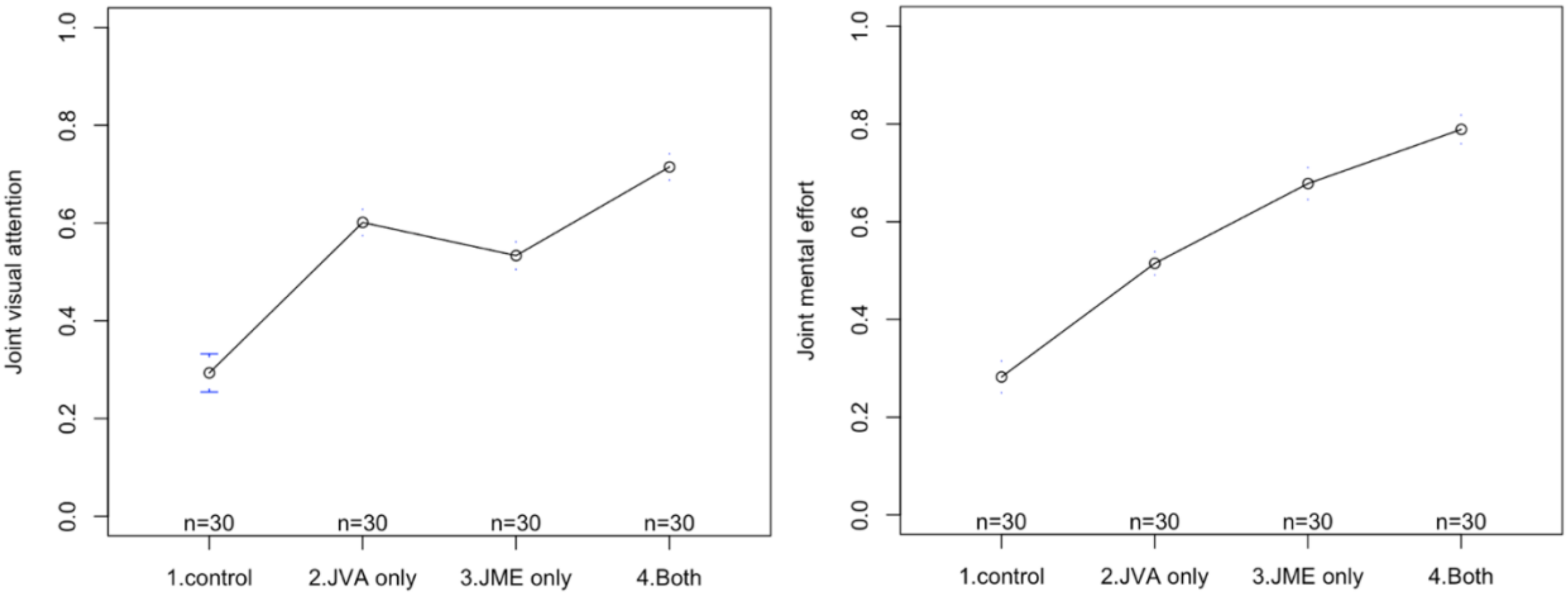

*Figure 10: JVA and JME for four reactive feedback conditions in the study 2.*

Next, we analyse the causal relation between JVA and JME for the different reactive feedback conditions. From figure 11, we observe that for "JME only" and "both" feedback conditions JME causes JVA (all the pairs from both feedback conditions are in the first quadrant). Moreover, for most of the pairs from "JVA only" feedback condition we can claim that JVA causes JME (26 out of 30 pairs from JVA only feedback condition are in the fourth quadrant while 4 out of 30 pairs from JVA only feedback condition are in the third quadrant). Finally, we observe that for control condition, only 10 out of 30 pairs are in the first quadrant and therefore we can claim, for thse pairs, that JME causes JVA but, for 20 out of 30 pairs are in the second and the third quadrants therefore, we can not claim any causality between JME and JVA for these pairs. The pairwise comparisons for the causality effect size and significance are shown in the table 2. From the pairwise comparison between the “JME only” and “both” feedback conditions, where JME causes JVA, the causality is significantly stronger for the “both” condition than the “JME only” condition.

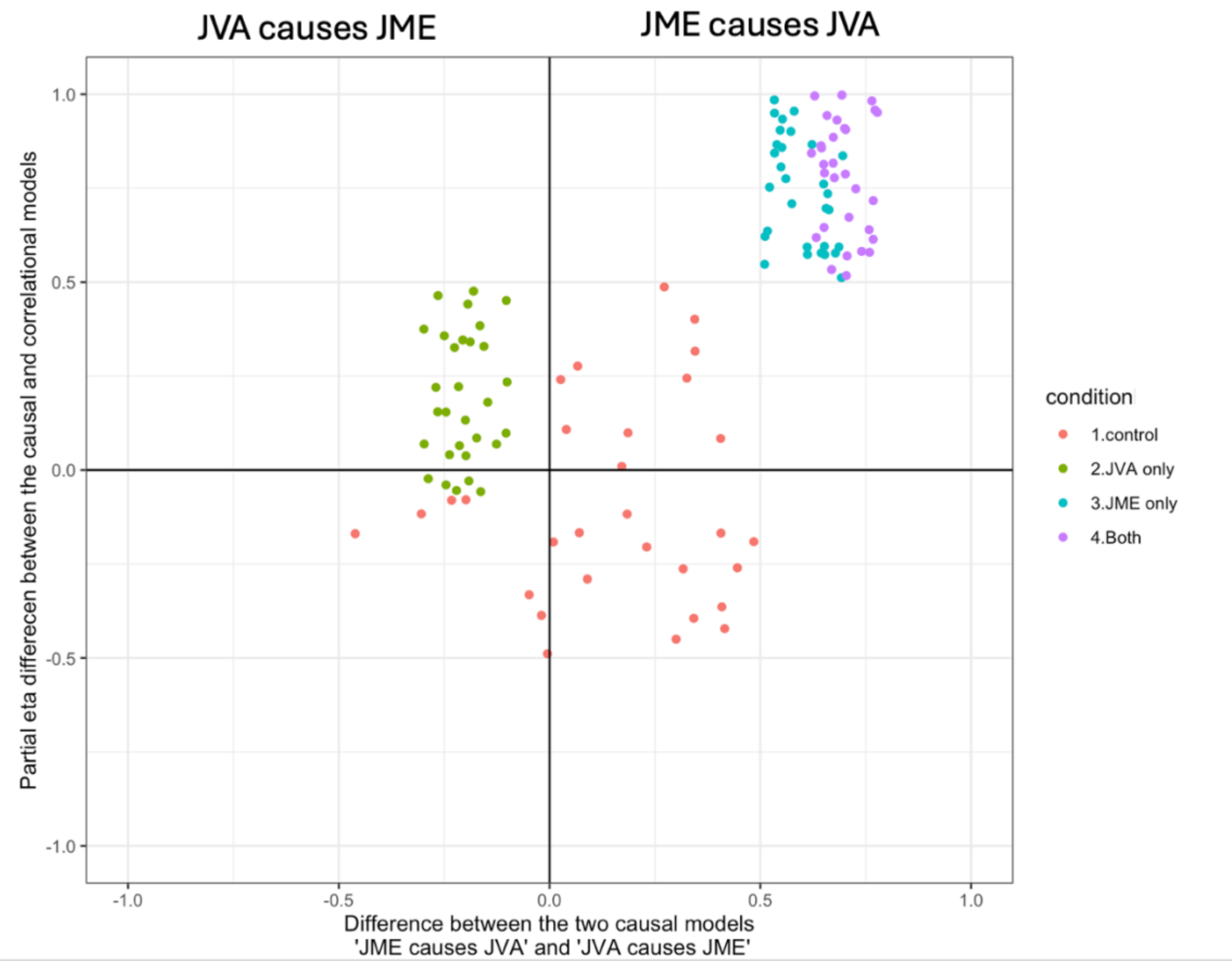

*Figure 11: causal analysis for the JVA and JME for four reactive feedback conditions in the study 2.*

*Table 2: pairwise ANOVA comparisons for causality effect size and significance for the four reactive feedback conditions (study 2).*

| | | Causality effect size | | Causality significance | |
|---|---|---|---|---|---|
| | | F | P | F | P |
| Control | JVA only | 62.71 | <.0001 | 25.05 | <.0001 |
| Control | JME only | 94.18 | <.0001 | 227.28 | <.0001 |
| Control | Both | 145.98 | <.0001 | 243.89 | <.0001 |
| JVA only | JME only | 273.5 | <.0001 | 177.22 | <.0001 |
| JVA only | Both | 442.0 | <.0001 | 195.77 | <.0001 |
| JME only | Both | 49.98 | <.0001 | 1.12 | 0.29 |

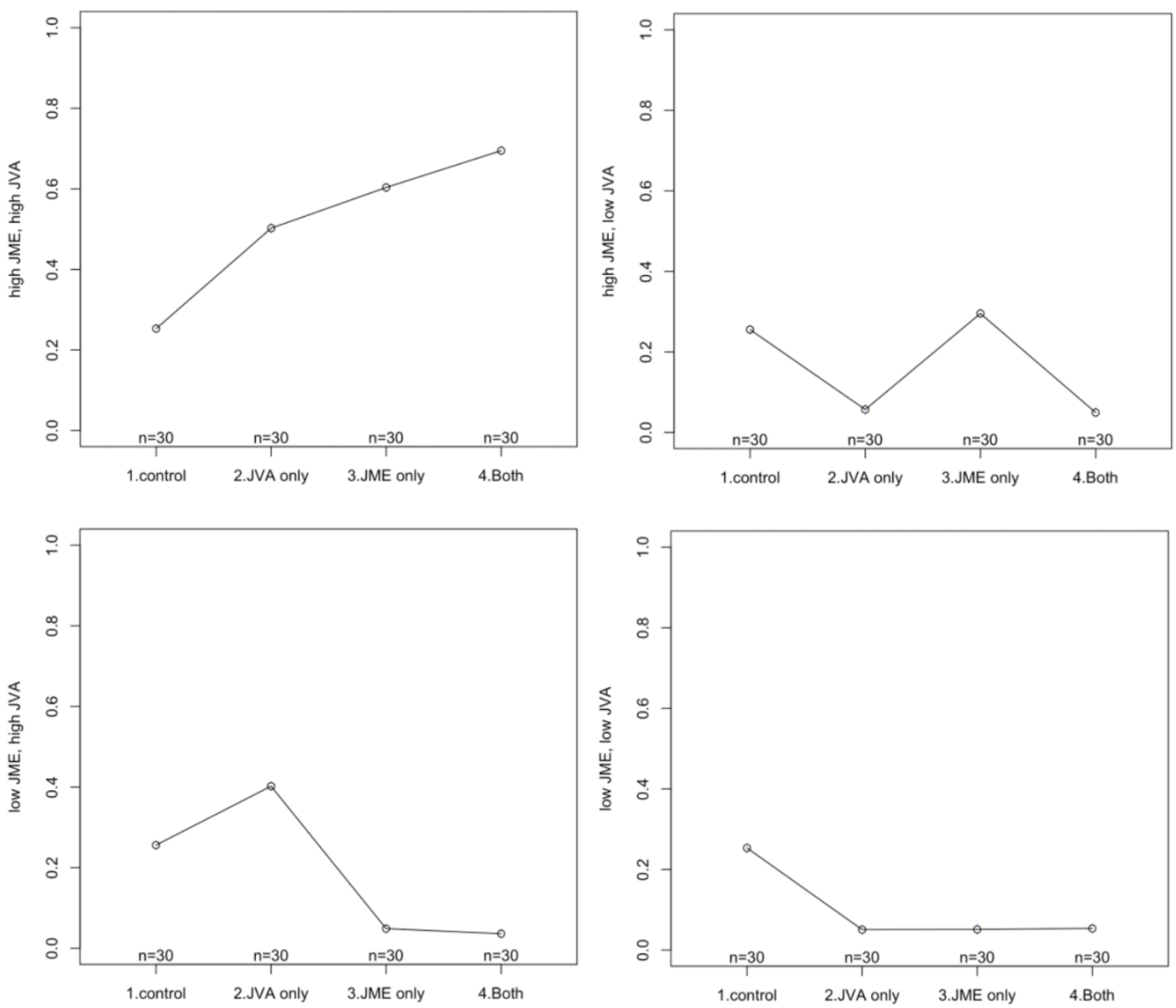

*Figure 12: proportions of the different JME-JVA episodes for the four reactive feedback conditions in the study 2.*

Finally, we analyse the proportions of JME JVA episodes across the four reactive feedback conditions. For "high JME high JVA" episodes, there is significant relation with the conditions ($F[3, 116] = 133.46$, $p < .0001$). Further pairwise comparisons (Table 3, columns 3 and 4, Figure 12 top-left) show that the control condition had the lowest proportion of "high JME high JVA" episodes, followed by "JVA only" and then by "JME only". Finally, "both" condition has the highest proportion of "high JME high JVA" episodes.

For the "high JME low JVA" episodes, there is significant relation with the conditions ($F[3, 116] = 61.03$, $p < .0001$). Further pairwise comparisons (Table 3, columns 5 and 6, Figure Figure 12 top-right) show that the "JVA only" and "both" conditions had the lowest proportion of "high JME low JVA" episodes (no significant difference between these two condition), followed by control condition. Finally, "JME only" condition has the highest proportion of "high JME low JVA" episodes.

For the "low JME high JVA" episodes, there is significant relation with the conditions (F[3, 116] = 122.92, p < .0001). Further pairwise comparisons (Table 3, columns 7 and 8, Figure 12 bottom-left) show that the "JME only" and "both" conditions had the lowest proportion of "low JME high JVA" episodes (no significant difference between these two condition), followed by control condition. Finally, "JVA only" condition has the highest proportion of "low JME high JVA" episodes.

For the "low JME low JVA" episodes, there is significant relation with the conditions (F[3, 116] = 124.11, p < .0001). Further pairwise comparisons (Table 3, columns 9 and 10, Figure 12 bottom-right) show that the "JVA only", "JME only" and "both" conditions had the lowest proportion of "low JME low JVA" episodes (no significant difference among these three condition). Finally, control condition has the highest proportion of "low JME low JVA" episodes.

*Table 3: pairwise ANOVA comparisons for the proportions of the JME-JVA episodes for the four reactive feedback conditions (study 2).*

| | | High JME High JVA | | High JME Low JVA | | Low JME High JVA | | Low JME Low JVA | |
|---|---|---|---|---|---|---|---|---|---|
| | | F | P | F | P | F | P | F | P |
| Control | JVA only | 112.5 | <.0001 | 372.64 | <.0001 | 346.17 | <.0001 | 695.28 | <.0001 |
| Control | JME only | 218.9 | <.0001 | 32.26 | <.0001 | 401.2 | <.0001 | 666.35 | <.0001 |
| Control | Both | 398.2 | <.0001 | 480.25 | <.0001 | 693.61 | <.0001 | 698.77 | <.0001 |
| JVA only | JME only | 56.91 | <.0001 | 528 | <.0001 | 93.6 | <.0001 | .003 | 0.95 |
| JVA only | Both | 705.58 | <.0001 | 0.94 | 0.33 | 720.4 | <.0001 | 0.15 | 0.69 |
| JME only | Both | 41.27 | <.0001 | 683.3 | <.0001 | 2.96 | .09 | 0.10 | 0.74 |

## Study 3: Proactive feedback

For the proactive feedback tool, there is a positive impact on the debugging success. The feedback condition improves the debugging success significantly than the control condition (F[1,24] = 182.45, p < .0001). Next, we analyse the relation between JVA and the proactive feedback condition. We observe that there is a significant improvement in the JVA for the feedback condition than the JVA for the control condition (F[1,24] = 283.46, p < .0001, Figure 13 left panel). Similarly, that there is a significant improvement in the JME for the feedback condition than the JVA for the control condition ($F[1,24] = 770.82$, $p < .0001$, Figure 13 right panel).

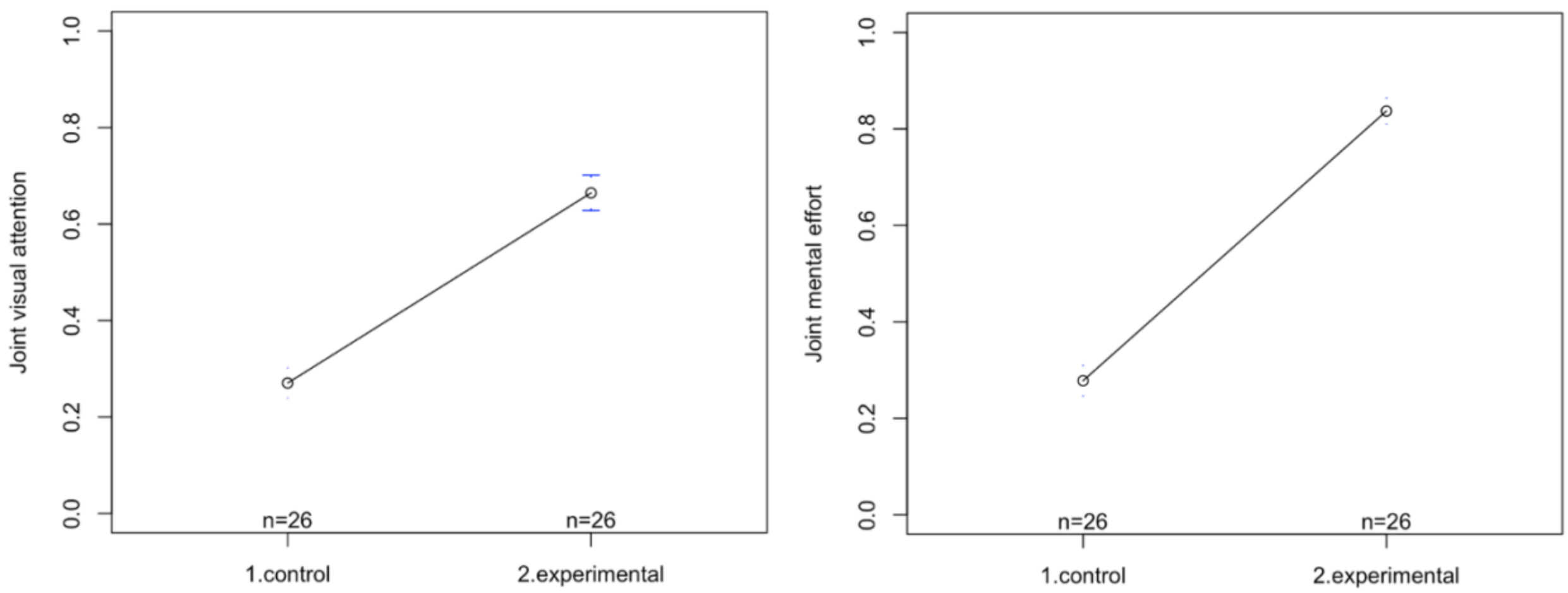

*Figure 13: JVA and JME for two proactive feedback conditions in the study 3.*

Next, we analyse the causal relation between JVA and JME for the proactive feedback conditions. From figure 14, we observe that for feedback conditions JME causes JVA (all the pairs from both feedback conditions are in the first quadrant). Finally, we observe that for control condition only 8 out of 26 pairs are in the first quadrant and therefore we can claim, for these pairs, that JME causes JVA but, for 18 out of 26 pairs are in the second and the third quadrants therefore, we can not claim any causality between JME and JVA for these pairs.

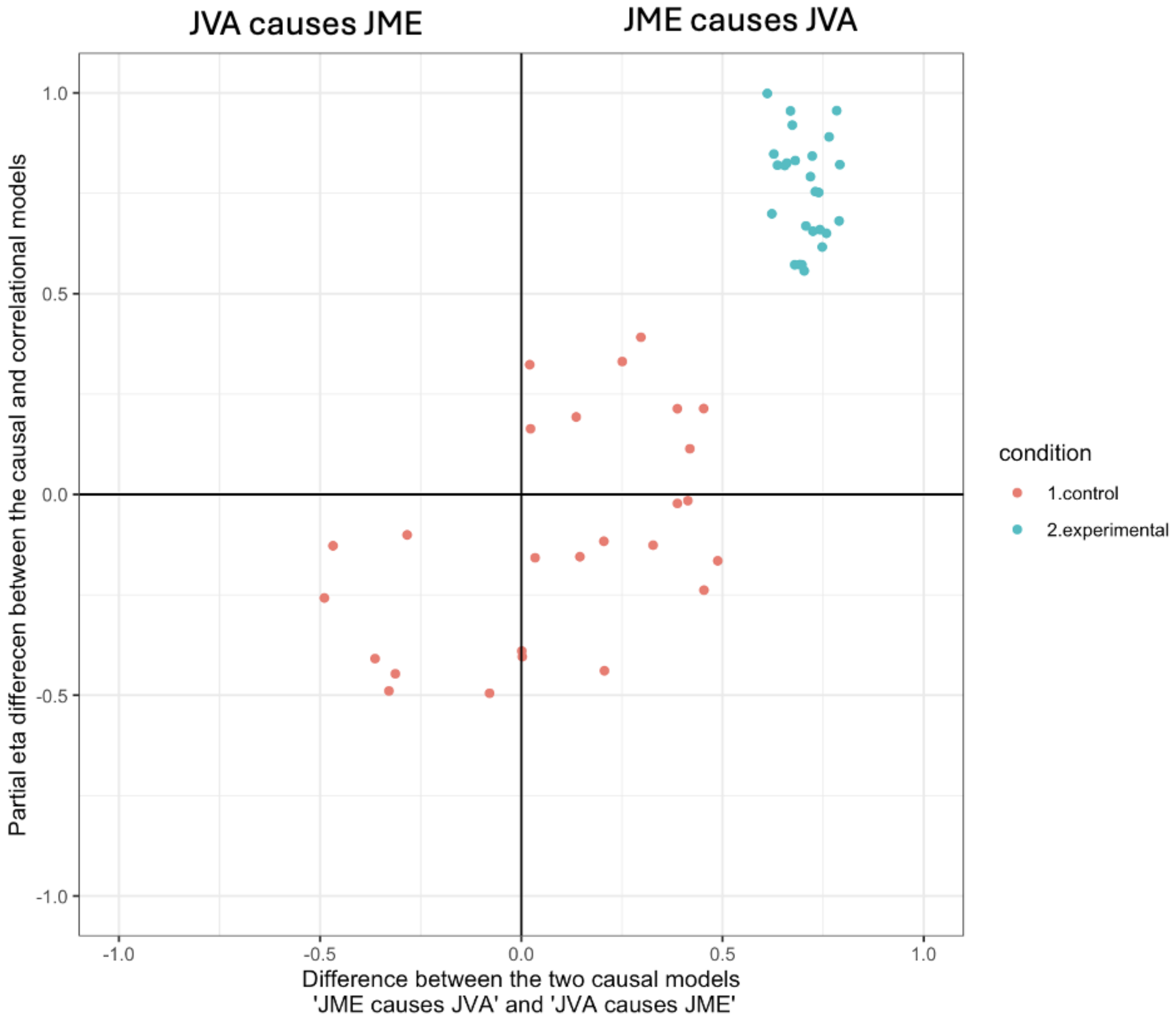

*Figure 14: causal analysis for the JVA and JME for two proactive feedback conditions in the study 3.*

Finally, we analyse the lengths of the JVA-JME episodes for proactive feedback conditions (Figure 15). We observe that the feedback condition had higher proportion of “high JME high JVA” episodes than the control condition ($F[1,26] = 2246.6$, $p < .0001$); while the control condition had higher proportions of “high JME low JVA” ($F[1,26] = 580.07$, $p < .0001$), “low JME high JVA”($F[1,26] = 565.08$, $p < .0001$), and “low JME low JVA” ($F[1,26] = 615.95$, $p < .0001$) episodes.

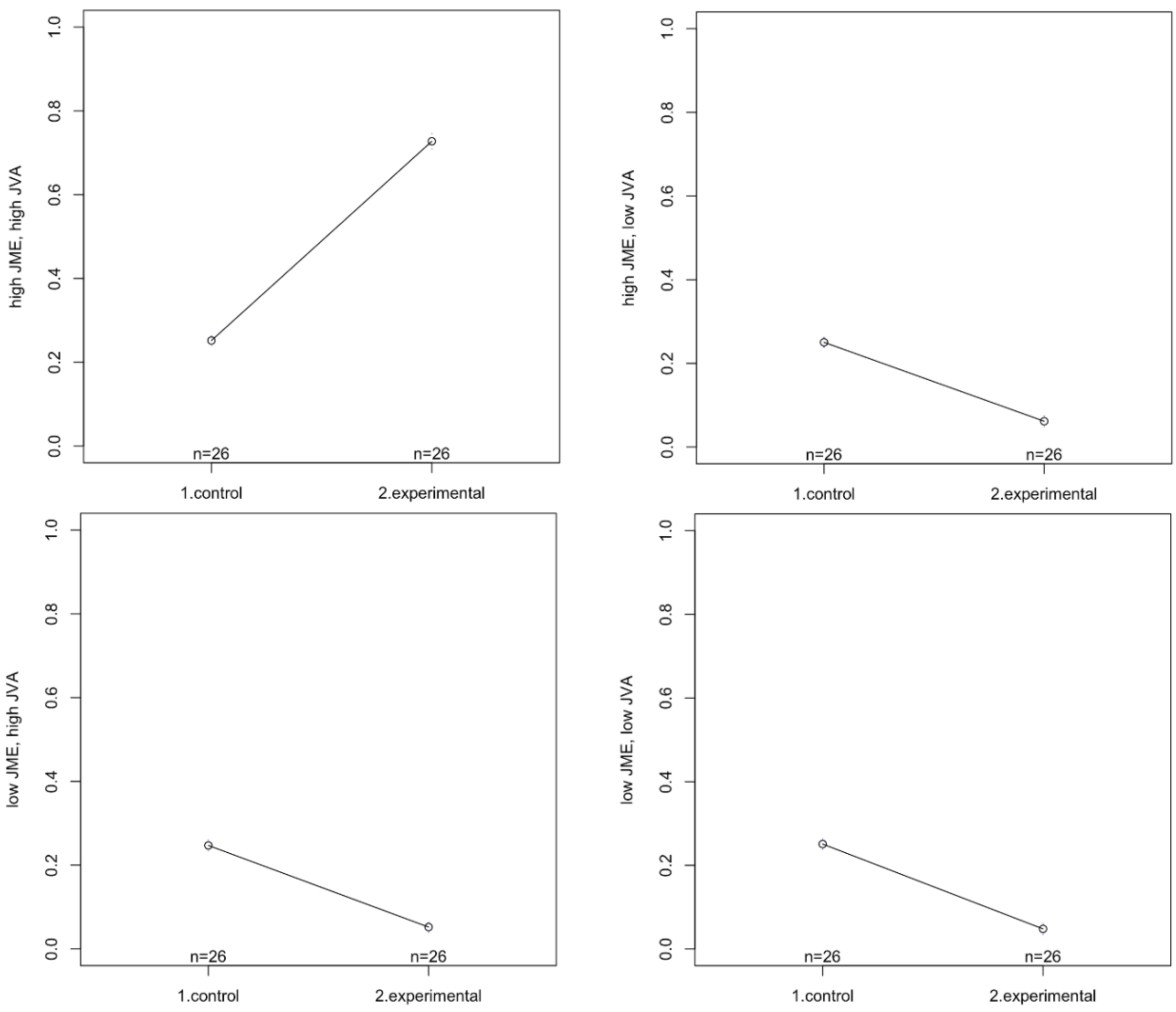

*Figure 15: proportions of the different JME-JVA episodes for the two proactive feedback conditions in the study 3.*

## Comparing the studies

For comparing the three studies, we chose the high performers from the first study, "both" condition from the reactive feedback study, and the experimental condition from the proactive feedback study. First, we analyse the differences among the studies with respect to the JVA and JME. We observe a significant difference among the three studies when JVA ($F[2, 37.67] = 77.65$, $p < .0001$) and JME ($F[2, 37.73] = 131.04$, $p < .0001$) are concerned. Further pairwise comparisons show that the JVA is the highest for "reactive combined feedback" followed by the JVA for "proactive experimental condition" and finally the JVA is the lowest for the "study 1 high performers" (Table 4, row 3, Figure 16 left panel). Similarly, pairwise comparisons show that the JME is the highest for "proactive experimental condition” followed by the JME for “reactive combined feedback” and finally the JME is the lowest for the “study 1 high performers” (Table 4, row 4, Figure 16 right panel).

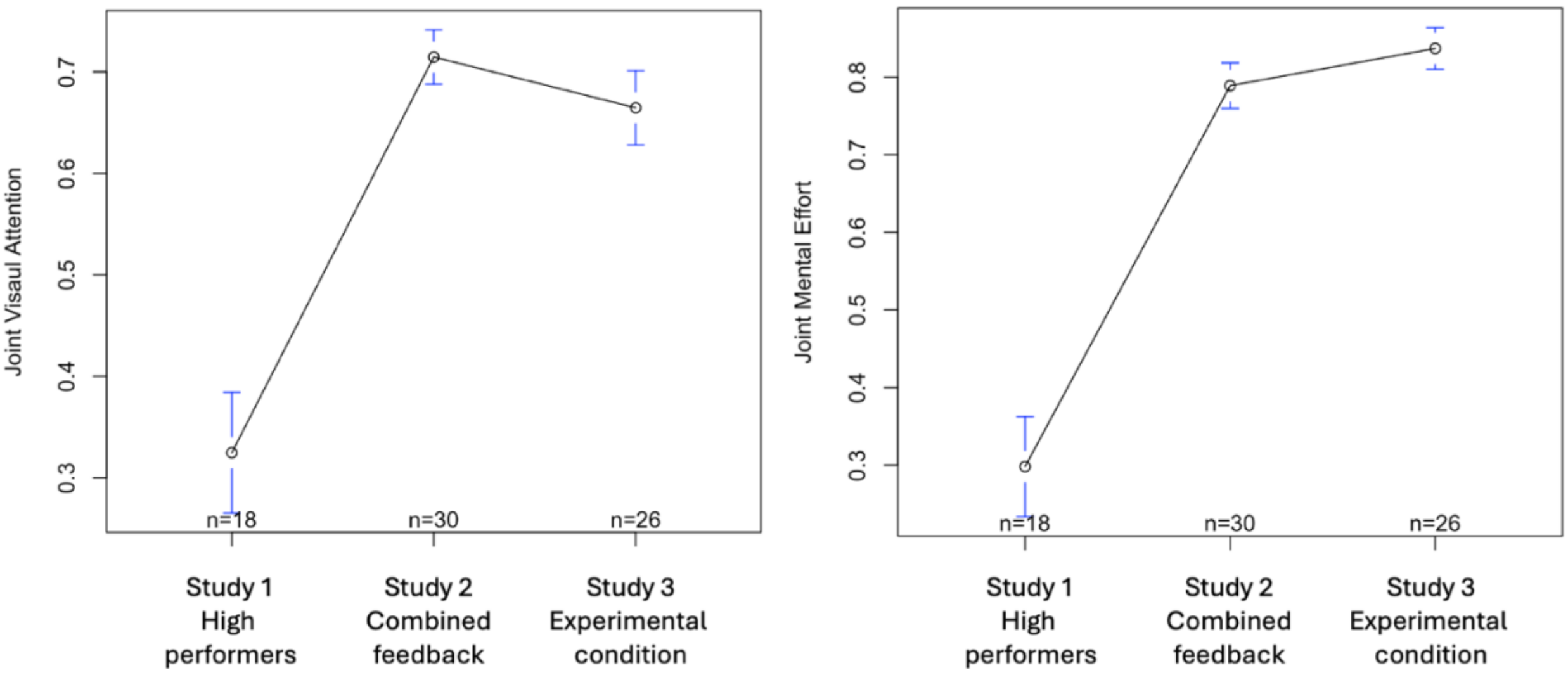

*Figure 16: JVA and JME from the subsets of the three studies for comparison.*

Next, we compare the studies considering the causal relation between JVA and JME. From figure 17, we observe that for all the three sets JME causes JVA but with different strengths. There is a significant difference among the three studies considering the effect size ($F[2, 34.25] = 9.79$, $p = .0004$) and significance ($F[2, 46.84] = 111.47$, $p < .0001$) of "JME causes JVA". Further pairwise comparisons show that there is no difference between “reactive combined feedback” and “proactive experimental condition” considering both effect size and significance of “JME causes JVA” (Table 4, rows 5 and 6, Figure 17). On the other hand, “study 1 high performance” has significantly lower effect size and significance for “JME causes JVA” than those for both “reactive combined feedback” and “proactive experimental condition” (Table 4, rows 5 and 6, Figure 17).

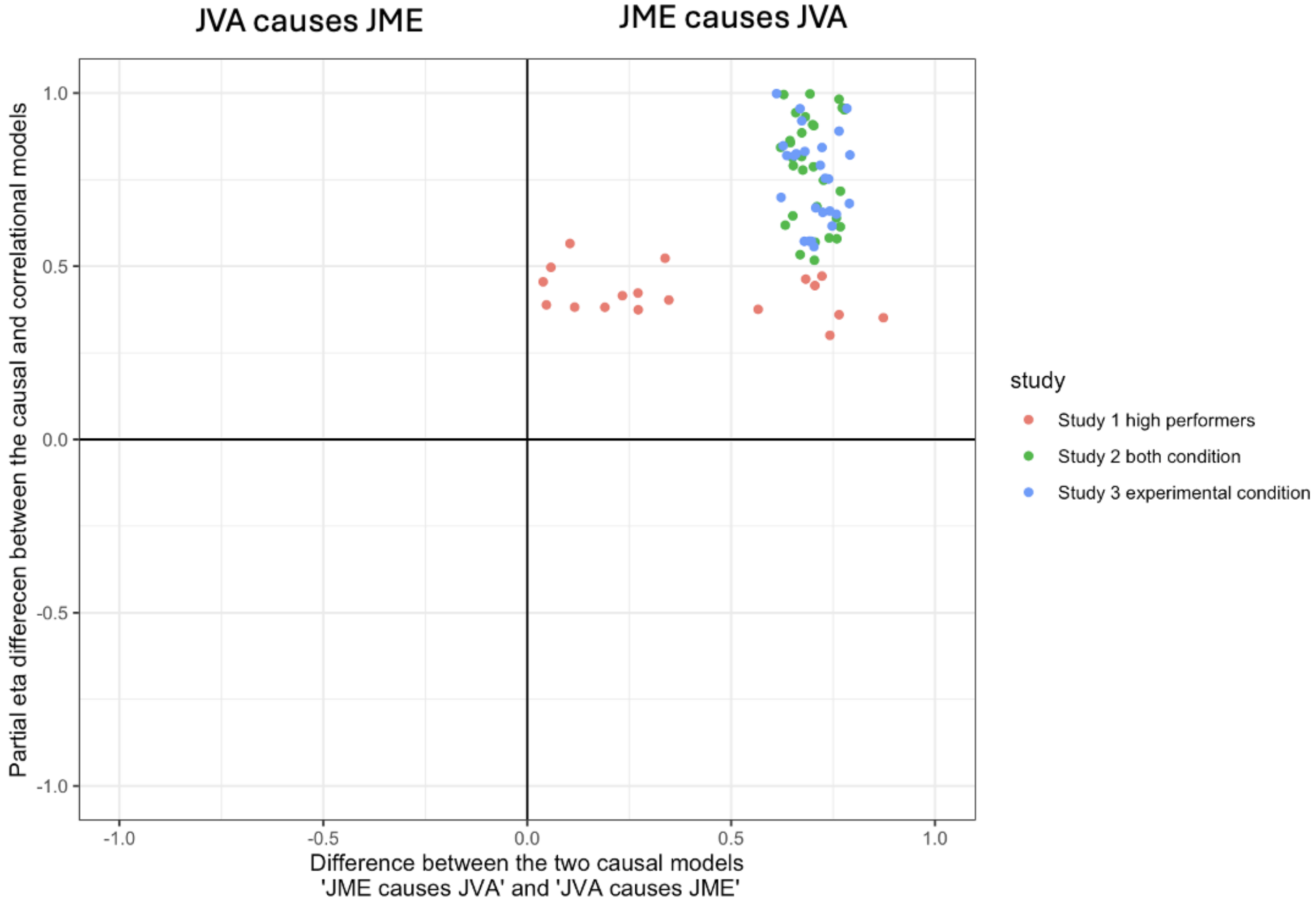

*Figure 17: causal analysis for the JVA and JME from the subsets of the three studies for comparison..*

Finally, we compare the proportions of the JVA JME episodes for the three studies. There is a significant difference between the three studies regarding the proportions of "high JME high JVA" episodes ($F[2, 33.43] = 40.63$, $p < .0001$). Further pairwise comparisons show that "proactive experimental condition" had the highest proportion of the "high JME high JVA" episodes, followed by that for the "reactive combined condition", and finally the "study 1 high performance" had the lowest proportion of the "high JME high JVA" episodes (Table 4, row 7, Figure 18 top-left). Second, there is also a significant difference between the three studies regarding the proportions of "high JME low JVA" episodes ($F[2, 43.89] = 248.61$, $p < .0001$). Further pairwise comparisons show that "study 1 high performance" had higher proportion of the "high JME low JVA" episodes, than that for "reactive combined condition" and "proactive experimental condition" (no significant difference between the last two datasets, Table 4, row 8, Figure 18 top-right). Third, there is also a significant difference between the three studies regarding the proportions of "low JME high JVA" episodes ($F[2,40.82] = 151.54$, $p < .0001$). Further pairwise comparisons show that "study 1 high performance" had higher proportion of the “low JME high JVA” episodes, than that for “reactive combined condition” and “proactive experimental condition” (no significant difference between the last two datasets, Table 4 row 9, Figure 18 bottom-left). Finally, there is also a significant difference between the three studies regarding the proportions of “low JME low JVA” episodes ($F[2, 46.95] = 9.21$, $p = .0004$). Further pairwise comparisons show that “study 1 high performance” had higher proportion of the “low JME low JVA” episodes, than that for “reactive combined condition” and “proactive experimental condition” (no significant difference between the last two datasets, Table 4, row 10, Figure 18 bottom-right).

*Table 4: pairwise ANOVA tests for various dependent variables from the subsets of the three studies for comparison.*

| | Study 1 VS Study 2 | | Study 1 VS Study 3 | | Study 2 VS Study 3 | |
|---|---|---|---|---|---|---|
| | F [df2, F-value] | P | F [df2, F-value] | P | F [df2, F-value] | p |
| JVA | 24.47, 157.24 | <.0001 | 29.94, 104.05 | <.0001 | 47.52, 5.11 | .02 |
| JME | 24.61, 211.57 | <.0001 | 23.24, 263.18 | <.0001 | 53.97, 6.14 | .01 |
| Causality Effect size | 17.56, 19.46 | .0003 | 17.53, 19.92 | .0003 | 50.43, 0.12 | .73 |
| Causality Significance | 42.96, 128.25 | <.0001 | 38.84, 154.6 | <.0001 | 53.99, 0.21 | .65 |
| High JME High JVA | 18.58, 66.35 | <.0001 | 21.21, 82.51 | <.0001 | 42.19, 9.90 | .003 |
| High JME Low JVA | 39.41, 395.49 | <.0001 | 40.08, 381.05 | <.0001 | 52.22, 0.43 | .51 |
| Low JME High JVA | 31.17, 298.02 | <.0001 | 35.15, 210.39 | <.0001 | 50.71, 3.43 | .06 |
| Low JME Low JVA | 45.55, 11.96 | .001 | 39.62, 12.37 | .001 | 52.62, 0.05 | .01 |

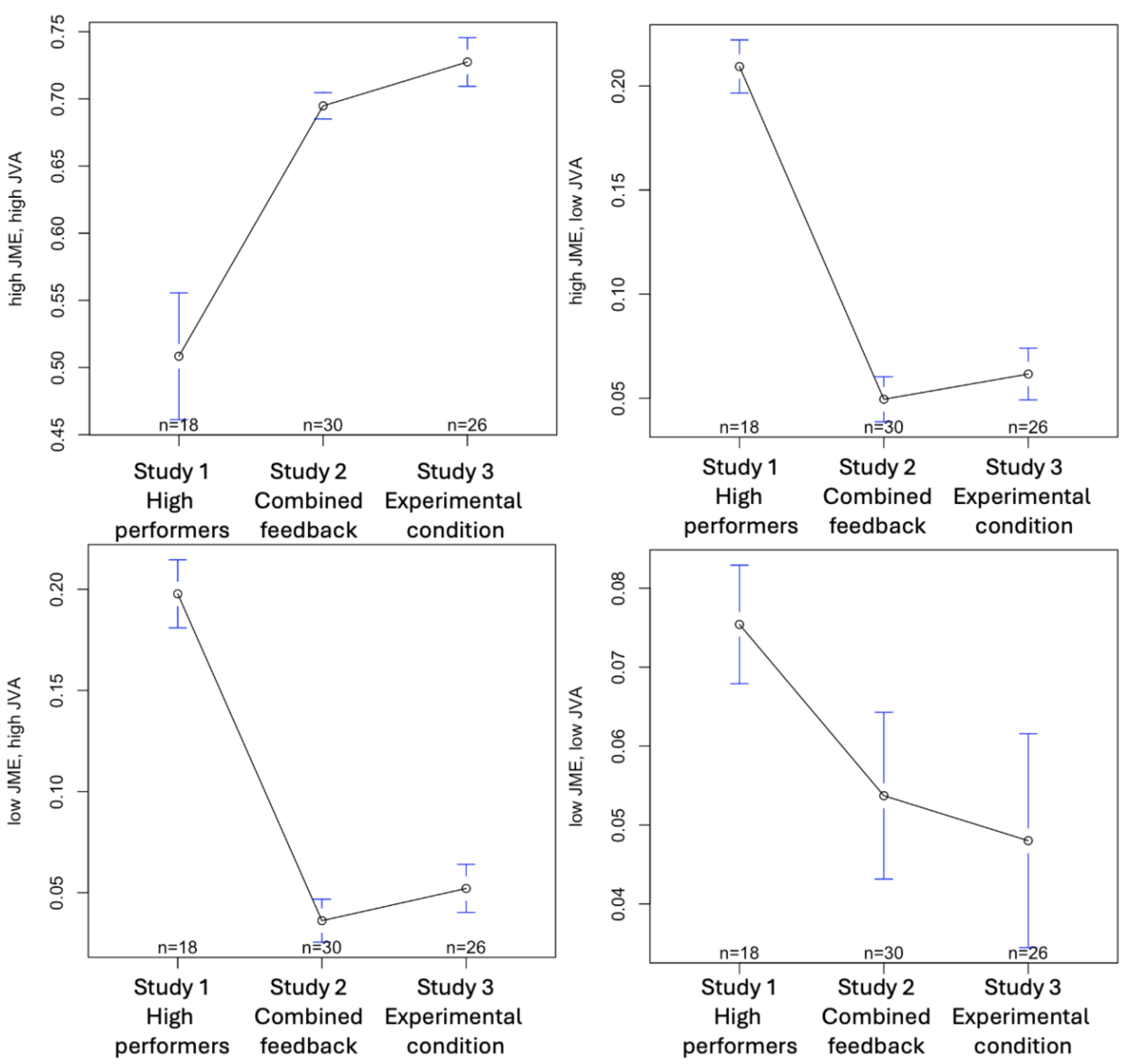

*Figure 18: proportions of the different JME-JVA episodes from the subsets of the three studies for comparison..*

# Discussion and Conclusions

This work advances CSCL research by providing a multi-study, process-oriented account of how socially shared regulation of learning (SSRL) unfolds in pair programming and how AI-driven adaptive support can strengthen these regulatory processes. Across three studies, the results converge on a central finding: productive collaboration is characterized by the alignment and coordination of joint mental effort (JME) and joint visual attention (JVA), rather than by either process alone. This alignment is not static, but dynamically regulated through time, shaped by task demands, interactional breakdowns, and the presence or absence of adaptive support.

## SSRL as a dynamic coordination of effort and attention

The findings across the three studies strongly support conceptualizing socially shared regulation of learning (SSRL) as a dynamic coordination of cognitive effort and attentional alignment, rather than as a static or purely behavioral phenomenon. From a theoretical standpoint, SSRL has been defined as a collective process through which group members jointly regulate cognition, motivation, emotion, and behavior toward shared goals (Järvelä & Hadwin, 2013; Panadero et al., 2015). The present results operationalize this definition by demonstrating how shared regulation manifests moment-to-moment through the coupling of joint mental effort (JME) and joint visual attention (JVA).

Across all studies, high-performing and well-supported dyads consistently exhibited a greater proportion of high-JME–high-JVA episodes, indicating sustained periods in which collaborators were simultaneously cognitively engaged and attentively aligned. These episodes can be interpreted as empirically observable instantiations of effective SSRL, during which groups successfully co-construct task understanding, monitor progress, and adjust strategies in response to emerging challenges. This interpretation aligns with prior work emphasizing that SSRL is inherently temporal and cyclical, involving iterative phases of planning, monitoring, and evaluation that unfold through interaction (Järvelä et al., 2019; Malmberg et al., 2015).

In contrast, low-performing and unsupported dyads were characterized by a dominance of mismatched episodes (e.g., high JME–low JVA or low JME–high JVA), reflecting breakdowns in shared regulation. These patterns suggest that individual engagement alone, whether cognitive or attentional, is insufficient to sustain productive collaboration. Instead, SSRL requires that effort and attention become mutually contingent across group members. This finding empirically substantiates theoretical claims that socially shared regulation cannot be reduced to parallel individual self-regulation processes, but instead emerges through transactive exchanges in which regulatory acts are jointly negotiated and enacted (Isohätälä et al., 2016; Sulla et al., 2023).

Crucially, the causal analyses deepen this interpretation by revealing that in successful collaboration, joint mental effort systematically precedes and predicts joint visual attention. This causal ordering suggests that shared cognitive engagement functions as a regulatory driver that organizes attention toward relevant task elements, rather than attention passively producing cognitive alignment. From an SSRL perspective, this supports the view that groups must first establish shared understanding, epistemic alignment, and strategic intent before attention can be productively coordinated around shared artifacts (Järvelä & Hadwin, 2013; Ito & Umemoto, 2022). In pair programming, this implies that looking at the same code is only meaningful when it is grounded in aligned interpretations of the problem and solution space.

This finding extends prior eye-tracking research in CSCL, which has repeatedly shown associations between joint visual attention and collaboration quality (Schneider et al., 2013; Bryant et al., 2019; Papadopoulos et al., 2017), by clarifying the regulatory mechanism underlying this relationship. While earlier studies often treated joint attention as a proxy for collaboration, the present results demonstrate that attention is embedded within a broader regulatory system, orchestrated by shared cognitive effort. This interpretation resonates with calls to move beyond surface indicators of coordination toward deeper models of collaborative regulation that integrate cognitive and metacognitive dimensions (Steier & Davidsen, 2021; Järvelä et al., 2014).

The modulation of JME–JVA dynamics through adaptive feedback further reinforces the view of SSRL as a malleable, design-sensitive process. Both reactive and proactive feedback systematically increased the prevalence of productive regulatory episodes and strengthened the causal coupling from JME to JVA. These effects illustrate how external regulation, implemented through AI-driven scaffolding, can temporarily support or restore socially shared regulation when groups struggle to sustain it independently. This aligns with prior work on co-regulation and socially shared regulation, which emphasizes that external supports can bootstrap regulatory competence until learners internalize more effective collaborative strategies (Hadwin et al., 2017; Järvelä et al., 2016).

Importantly, the differential causal patterns observed across feedback types also highlight a key theoretical distinction within SSRL. Conditions emphasizing attention cues alone sometimes reversed the causal direction (JVA causing JME), suggesting that externally imposed attentional alignment can prompt cognitive engagement, but may risk shallow coordination if not accompanied by deeper regulatory alignment. This finding echoes concerns in the SSRL literature that regulation driven primarily by surface-level coordination may not translate into sustained learning gains unless cognitive and metacognitive processes are jointly regulated (Malmberg et al., 2015; Panadero et al., 2015).

Taken together, these results position SSRL as a dynamic coordination system in which shared cognitive effort organizes attentional focus, and attention, in turn, stabilizes and sustains collective engagement with shared artifacts. By empirically linking SSRL theory to measurable, time-sensitive indicators, this work advances both the conceptual clarity and methodological tractability of socially shared regulation in CSCL.

## Cognitive alignment as a driver of attentional coordination

The results of this work provide strong empirical support for conceptualizing cognitive alignment as a primary driver of attentional coordination in collaborative learning. While prior CSCL research has repeatedly demonstrated associations between joint visual attention and collaboration quality, the present findings go beyond correlation by revealing a consistent causal ordering in which joint mental effort (JME) precedes and predicts joint visual attention (JVA) in successful collaboration. This causal structure has important theoretical implications for how attentional coordination should be understood within socially shared regulation of learning (SSRL).

From a theoretical perspective, cognitive alignment refers to the degree to which collaborators share task understanding, epistemic goals, and problem representations (Barron, 2003; Dillenbourg, 1999). SSRL theory posits that such alignment is a prerequisite for effective regulation, enabling groups to coordinate strategies, monitor progress, and respond adaptively to challenges (Järvelä & Hadwin, 2013; Panadero et al., 2015). The present results empirically substantiate this claim by demonstrating that when collaborators' mental effort is synchronized, reflecting aligned cognitive engagement, attention subsequently converges on shared task-relevant elements.

In Study 1, all high-performing dyads exhibited a stable causal pattern in which JME Granger-caused JVA, whereas low-performing dyads showed no consistent causal relationship between the two processes. This distinction suggests that attentional alignment is not merely a behavioral precursor to cognitive coordination, but rather an outcome of successful cognitive alignment. In low-performing dyads, attention may be fragmented, reactive, or driven by individual exploration rather than by shared understanding, leading to unstable or non-predictive relations between effort and attention. This interpretation aligns with prior qualitative and quantitative studies showing that simply looking at the same object does not guarantee shared meaning or productive collaboration (Lämsä et al., 2022; Sangin et al., n.d.).

These findings refine earlier eye-tracking work that treated joint visual attention as a proxy for collaboration quality (Schneider et al., 2013; Papadopoulos et al., 2017; Bryant et al., 2019). While joint attention remains an important indicator, the present results suggest that its pedagogical significance depends on whether it is grounded in cognitive alignment. Without shared understanding, attentional synchrony may reflect superficial coordination or even social compliance, rather than meaningful engagement with the task. This distinction echoes critiques in the CSCL literature cautioning against equating behavioral synchrony with deep collaboration (Dillenbourg et al., 2009; Steier & Davidsen, 2021).

The causal role of cognitive alignment becomes even clearer in the adaptive feedback studies. In Study 2, feedback conditions that targeted mental effort, either alone or in combination with attention cues, consistently preserved the theoretically grounded direction of causality (JME causing JVA). In contrast, the JVA-only condition often reversed this relationship, producing cases where JVA appeared to cause JME. This pattern suggests that externally induced attentional alignment can stimulate cognitive engagement, but does not reliably establish the deep alignment necessary for sustained shared regulation. Such findings resonate with research on awareness tools, which shows that gaze or cursor sharing can improve coordination but may also introduce distraction or shallow coupling if not embedded in broader regulatory support (Cheng et al., 2021; Hayashi, 2020).

Study 3 further strengthens this interpretation by showing that proactive, forecast-based feedback yields the strongest and most consistent causal coupling from JME to JVA. By anticipating future breakdowns in cognitive alignment, the system was able to intervene before attentional fragmentation emerged, thereby preserving continuity in shared focus. This finding aligns with SSRL models that emphasize the anticipatory and proactive nature of effective regulation, where groups adjust strategies and engagement in advance of overt failure (Järvelä et al., 2019; Hadwin et al., 2017). It also highlights the potential of predictive analytics to support regulation without imposing rigid scripts or undermining learner agency.

From a design perspective, these results suggest that attention should be treated as a regulated outcome rather than a primary control variable in CSCL systems. While making attention visible can be useful, effective collaborative support must prioritize mechanisms that foster shared understanding, balanced cognitive effort, and epistemic alignment. This perspective challenges designs that rely solely on gaze awareness or visual coupling and instead advocates for multi-dimensional support that integrates cognitive, metacognitive, and attentional processes (Malmberg et al., 2015; Silva et al., 2023).

Taken together, the findings position cognitive alignment as the organizing force that structures attentional coordination in collaborative learning. By empirically demonstrating the causal primacy of shared mental effort, this work advances theoretical models of SSRL and provides concrete guidance for the design of AI-supported CSCL environments. Attention, in this view, is not merely where collaborators look together, but where shared cognition has already brought them.

## Effects of reactive adaptive feedback on regulatory processes

The results of Study 2 demonstrate that reactive, real-time adaptive feedback can substantially reshape collaborative regulatory processes by influencing how cognitive effort and attention are coordinated over time. Across all feedback conditions, dyads receiving adaptive support significantly outperformed the control condition, confirming prior findings that adaptive collaborative learning support (ACLS) can enhance both collaboration quality and learning outcomes (VanLehn, 2016; Suebnukarn & Haddawy, 2006). However, the present results extend this literature by revealing how different types of feedback differentially affect the underlying dynamics of socially shared regulation of learning (SSRL).

Feedback conditions targeting both joint mental effort (JME) and joint visual attention (JVA) produced the strongest improvements in performance, process measures, and regulatory coherence. Dyads in the combined condition exhibited the highest levels of JME and JVA, the strongest causal coupling from JME to JVA, and the greatest proportion of productive high-JME–high-JVA episodes. These findings suggest that supporting multiple regulatory dimensions simultaneously enables groups to maintain cognitive alignment while sustaining shared focus, a core requirement for effective SSRL (Järvelä & Hadwin, 2013; Panadero et al., 2015).

In contrast, single-channel feedback conditions revealed important nuances. JME-only feedback improved performance and preserved the theoretically grounded causal pattern in which cognitive alignment precedes attentional coordination. This indicates that interventions targeting cognitive engagement can indirectly stabilize attention by restoring balance in mental effort across collaborators. Such findings align with SSRL research emphasizing that shared understanding and epistemic alignment are foundational for coordinated regulation (Isohätälä et al., 2016; Ito & Umemoto, 2022).

JVA-only feedback, while effective in increasing shared attention, often altered the causal structure of regulation, with attention preceding mental effort for many dyads. This pattern suggests that externally induced attentional alignment can prompt engagement, but may not reliably foster deep cognitive alignment. Prior work on gaze awareness tools similarly shows that while attention cues can enhance coordination, they may also lead to superficial coupling or distraction if not embedded in broader regulatory support (Cheng et al., 2021; Lämsä et al., 2022). From an SSRL perspective, such patterns reflect regulation driven by surface-level coordination rather than jointly constructed understanding.

The episode-level analyses further clarify these effects. Combined feedback significantly increased the prevalence of high-JME–high-JVA episodes while reducing mismatched or unproductive states. This redistribution of regulatory episodes illustrates how reactive feedback can act as a stabilizing mechanism, helping groups recover from momentary breakdowns in shared regulation without prescribing rigid interaction scripts. These findings resonate with models of co-regulation, where external scaffolding temporarily supports regulation until learners regain control (Hadwin et al., 2017).

## Proactive feedback and predictive regulation support

The results of Study 3 demonstrate that proactive, forecast-based adaptive feedback offers distinct advantages over reactive support by enabling early intervention in collaborative regulatory processes. Dyads receiving proactive feedback significantly outperformed the control condition and exhibited higher levels of joint mental effort (JME), joint visual attention (JVA), stronger causal coupling from JME to JVA, and the highest proportion of productive high-JME–high-JVA episodes. These findings suggest that anticipating future breakdowns in shared regulation is more effective than responding after misalignment has already emerged.

From a theoretical perspective, proactive feedback aligns closely with models of socially shared regulation of learning (SSRL) that emphasize the anticipatory and cyclical nature of regulation. Effective groups do not merely react to difficulties but continuously monitor emerging conditions and adjust strategies in advance (Järvelä & Hadwin, 2013; Järvelä et al., 2019). By forecasting collaboration states 30 seconds ahead, the system operationalized this anticipatory function, enabling external support to scaffold regulation at moments when learners were most vulnerable to breakdowns in cognitive alignment or attentional coordination.

The observed strengthening of the causal relationship in which JME causes JVA under proactive feedback is particularly noteworthy. Compared to both the control condition and high-performing dyads in the unsupported study (study 1), proactive feedback yielded the most stable and pronounced cognitive-to-attentional causal structure. This suggests that predictive support not only enhances performance but also reinforces the theoretically grounded ordering of regulatory processes, helping groups sustain shared understanding as the organizing force behind attentional coordination. Such findings extend prior work on adaptive collaborative learning support by demonstrating that prediction-based interventions can shape the quality of regulation, not just its outcomes (VanLehn, 2016; Neto et al., 2022).

Importantly, proactive feedback functioned as a form of intelligence augmentation rather than automation. Interventions were triggered selectively and remained lightweight, preserving learners' agency and responsibility for regulation. This design choice addresses long-standing concerns in CSCL about over-automation and the risk of imposing rigid interaction patterns that undermine collaborative autonomy (Rummel et al., 2016; Roschelle, 2021). Instead, the system acted as a co-regulator, supporting learners' capacity to maintain shared regulation without dictating how collaboration should unfold.

The episode-level findings further reinforce the value of predictive support. Proactive feedback produced the highest proportion of sustained high-JME–high-JVA episodes while minimizing time spent in mismatched or low-regulation states. This suggests that early intervention helps preserve continuity in collaboration, preventing minor misalignments from escalating into prolonged breakdowns. Such continuity is especially important in complex tasks like pair programming, where interruptions in shared understanding can quickly derail progress.

## Broader implications for CSCL and learning analytics

Collectively, these findings contribute to ongoing efforts to move CSCL research beyond outcome-focused evaluation toward a deeper understanding of collaborative processes. Methodologically, the integration of dual eye-tracking, pupillometry, episode-based modeling, and causal inference provides a powerful framework for studying SSRL as a dynamic phenomenon. This addresses long-standing challenges related to ecological validity and process sensitivity in multimodal learning analytics, particularly in authentic, high-variability collaborative tasks.

Conceptually, the results reinforce a shift from designing for structure alone toward designing for regulation. AI systems, when grounded in learning theory, can support learners' capacity to regulate cognition, attention, and effort together, rather than optimizing isolated performance metrics. This perspective positions AI as a partner in collaboration, supporting shared responsibility and agency, which are the core principles of CSCL.

## Limitations and Future work

Despite its contributions, this work has several limitations that point to important directions for future research. First, all studies were conducted in controlled laboratory settings using pair programming tasks of comparable difficulty. While this allowed for precise measurement and causal analysis, future work should examine the robustness of these findings in more ecologically valid settings, such as classrooms, online courses, or longer-term collaborative projects where task structures and social dynamics are more diverse.

Second, the focus on dyads limits the generalizability of the findings to larger groups. SSRL processes become increasingly complex as group size grows, raising challenges for measuring and supporting shared regulation at scale. Future research should explore how JVA–JME indicators can be aggregated or adapted for triads or small groups, and how feedback visualizations can be designed to avoid information overload in multiparty collaboration.

Third, while JVA and JME capture critical aspects of shared regulation, they do not directly measure motivational, emotional, or discursive processes that are central to SSRL. Integrating additional data sources, such as speech, facial expressions, or self-reports, could provide a more holistic account of how cognitive, affective, and motivational regulation interact during collaboration.

Finally, although the proactive system demonstrated strong benefits, its forecasting models were trained and evaluated within a specific task domain. Future work should investigate the transferability of these models across tasks, domains, and learner populations, as well as explore how learners perceive and appropriate such adaptive support over time.

# Conclusions

This paper contributes to CSCL research by demonstrating that effective collaboration is grounded in the dynamic alignment of cognitive effort and attention and that these processes can be meaningfully supported through AI-driven adaptive feedback. Across three studies, we showed that high-performing dyads exhibit stronger joint visual attention, greater synchronization of mental effort, and a stable causal relationship in which cognitive alignment precedes attentional coordination. Reactive and proactive feedback further enhanced these regulatory processes, with proactive, forecast-based support yielding the most robust outcomes.

By grounding adaptive support in socially shared regulation of learning, this work advances a view of AI as an intelligence-augmenting partner that supports learners' capacity to regulate together rather than directing collaboration through rigid scripts. Methodologically, the proposed framework offers a scalable approach for studying and supporting SSRL in real time, addressing key challenges in learning analytics and adaptive systems.

Ultimately, these findings underscore the importance of designing CSCL environments that support not only what learners do together, but how they regulate their shared cognition, attention, and effort over time.

# Declarations

## Funding

The authors did not receive support from any organization for the submitted work

## Data and/or Code availability

Due to the ethical and privacy requirements, the original data cannot be shared. The measurements can be shared upon contacting the corresponding author.

## Competing interests

Author KS is on the editorial board of the journal.

## Author contributions

AG: conceptualization; design of the studies; data collection for study 2 and 3; verification and validation; writing and revising.

KS: conceptualization; design of the studies; data collection for study 1; data analysis; supervision; writing and revising.

# Appendix A

Table A1: Feedback combinations for different states.

A1 → Do nothing (positive feedback); A2 → Offer help through enabling GitHub Copilot; A3 → Offer help through enabling the gaze-awareness tool; A4 → Prompt to initiate dialogue; A5 → Prompt a task-based hint.

H = high, L = low, Avg = average

| SCENARIO | JVA | JME | ME1 | ME2 | A1 | A2 | A3 | A4 | A5 |
|---|---|---|---|---|---|---|---|---|---|
| 1 | H | H | H | H | | √ | | | √ |
| 2 | H | H | Avg | Avg | √ | | | | |
| 3 | H | H | L | L | | √ | | | |
| 4 | H | L | H | H | | √ | | √ | √ |
| 5 | H | L | H | L | | | | √ | |
| 6 | H | L | L | L | | | | √ | |
| 7 | H | L | Avg | H | | | | √ | |
| 8 | H | L | L | L | | | | √ | |
| 9 | H | L | L | L | | | | √ | |
| 10 | H | L | H | H | | | | √ | |
| 11 | H | L | L | L | | √ | | √ | |
| 12 | L | H | L | L | | √ | √ | | √ |
| 13 | L | H | L | L | | | √ | | |
| 14 | L | H | L | L | | √ | √ | | |
| 15 | L | L | H | H | | √ | √ | √ | √ |
| 16 | L | L | H | Avg | | | √ | √ | |
| 17 | L | L | H | L | | √ | √ | √ | |
| 18 | L | L | Avg | H | | | √ | √ | |
| 19 | L | L | Avg | Avg | | | √ | √ | |
| 20 | L | L | L | H | | √ | √ | √ | |
| 21 | L | L | L | L | | | √ | √ | |
| 22 | L | L | L | L | | √ | √ | √ | |

## Detailed Scenario Logic and Feedback Combinations

Table A1 defines 22 scenarios derived from combinations of Joint Visual Attention (JVA), Joint mental-effort (JME), and individual mental-efforts (ME1, ME2). Each configuration reflects a distinct collaboration state and triggers specific feedback mechanisms (A1–A5) in the Oculii system. Desired states correspond to high JVA and JME with average individual MEs, whereas suboptimal configurations invoke corrective or assistive feedback.

## Scenario description

- High JVA – High JME (Scenarios 1–3): Optimal engagement states; use of A1 ("do nothing") or A2/A5 for cognitive rebalancing.
    - These scenarios indicate strong shared attention and engagement. However, only Scenario 2 constitutes a desired state, where both individual MEs are balanced. Here, the system applies A1 (Do nothing) to preserve the optimal configuration. In Scenarios 1 and 3, the individual MEs are both too high or too low, respectively. To guide the pair toward the optimal average ME level, A2 (GitHub Copilot) is activated either to aid in discovering new solutions under excessive cognitive load or to accelerate task progress when the task is too easy. If the high-ME condition persists (Scenario 1 ), a corrective hint (A5) is subsequently provided to re-stabilize the pair's cognitive alignment.
- High JVA – Low JME (Scenarios 4–11): Visual alignment but cognitive divergence; initiate A4 (dialog) and optionally A2/A5.
    - Here, both partners remain visually coordinated, but their cognitive engagement diverges. The system prioritizes A4 (Dialog Prompt) to stimulate discussion and restore shared understanding. Dialogic reflection can help align individual effort levels, transforming low JME into a higher, more synchronized state. When both MEs are extreme, either high (Scenario 4 ) or low (Scenario 11) additional aids such as A2 (Copilot) and, where necessary, A5 (Hint) are enabled to regulate cognitive load and re-establish balance across the pair.
- Low JVA – High JME (Scenarios 12–14): Productive division of labour; maintain A3 (gaze-awareness) and support with A2 as needed.
    - These scenarios represent desired states of effective Division of Labour (DoL), either task-based or role-based. Although the pair is productively engaged, low JVA may hinder coordination. To support smooth transitions between sub-tasks, A3 (Gaze-awareness tool) is activated, allowing partners to re-synchronize their visual focus when needed. In Scenarios 12 and 14, A2 (Copilot) aids code exploration or rapid navigation. For Scenario 14, which may indicate either deep focus or disengagement, A6 (Question Prompt) is occasionally employed to assess the collaboration state and help re-engage the pair when necessary.
- Low JVA – Low JME (Scenarios 15–22): Disengaged or uncoordinated states; combine A3, A4, and when required A2/A5 to restore synchrony.

- Low levels of both indicators suggest disconnection or reduced engagement. Consequently, A3 (Gaze-awareness) and A4 (Dialog Prompt) are deployed to rebuild mutual focus and stimulate discussion. For configurations where both individual Mes deviate from average (Scenarios 15, 17, 20, 22 ), A2 (Copilot) provides external scaffolding to encourage progress. In Scenario 15, sustained high MEs trigger an additional A5 (Hint) to alleviate overload.